\documentclass[11pt]{article}

\usepackage[totalwidth=460truept,totalheight=605truept]{geometry}
\usepackage{amsfonts,latexsym,amssymb,amsmath,graphicx,accents,eucal,slashed,subfigure}
\usepackage{dsfont}
\usepackage[T1]{fontenc}

\global\arraycolsep=1truept

\numberwithin{equation}{section}

\usepackage{latexsym}
\usepackage[hidelinks]{hyperref}

\begin{document}

\bigskip \phantom{C}

\vskip1.4truecm

\begin{center}
{\huge \textbf{Aspects Of Perturbative Unitarity}}

\vskip 1truecm

\textsl{Damiano Anselmi}

\vskip .2truecm

\textit{Dipartimento di Fisica ``Enrico Fermi'', Universit\`{a} di Pisa, }

\textit{Largo B. Pontecorvo 3, 56127 Pisa, Italy}

\textit{and INFN, Sezione di Pisa,}

\textit{Largo B. Pontecorvo 3, 56127 Pisa, Italy}

\vskip.2truecm

damiano.anselmi@unipi.it

\vskip 1.5truecm

\textbf{Abstract}
\end{center}

We reconsider perturbative unitarity in quantum field theory and upgrade
several arguments and results. The minimum assumptions that lead to the
largest time equation, the cutting equations and the unitarity equation are
identified. Using this knowledge and a special gauge, we give a new, simpler
proof of perturbative unitarity in gauge theories and generalize it to
quantum gravity, in four and higher dimensions. The special gauge
interpolates between the Feynman gauge and the Coulomb gauge without double
poles. When the Coulomb limit is approached, the unphysical particles drop
out of the cuts and the cutting equations are consistently projected onto
the physical subspace. The proof does not extend to nonlocal quantum field
theories of gauge fields and gravity, whose unitarity remains uncertain.

\vfill\eject

\section{Introduction}

\label{s0}

\setcounter{equation}{0}

The problem of quantum gravity is the apparent incompatibility between
unitarity and renormalizability. For example, the quantization of Einstein
gravity gives a theory that is unitary, but not renormalizable \cite%
{thooftveltman,wein}. If the counterterms generated by renormalization are
included, the theory becomes renormalizable with infinitely many independent
couplings\footnote{%
For this reason, \textquotedblleft nonrenormalizable\textquotedblright\ and
\textquotedblleft renormalizable with infinitely many independent
couplings\textquotedblright\ are often used interchangeably.} and predictive
at low energies (see, for example, \cite{newgrav}). It is also possible to
build theories of quantum gravity that are renormalizable (with finitely
many independent couplings), but not unitary. One way to achieve this goal
is by including higher-derivative terms that make the propagators fall off
more rapidly at high energies \cite{stelle}.\ It is not known how to build a
theory that is renormalizable and unitary at the same time.

Perturbative unitarity is thus a key issue in quantum field theory. In
scalar and fermion theories it can be proved by means of the cutting
equations \cite{cutkosky,veltman}. In gauge theories, additional aspects
need to be addressed, such as the compensation between the Faddeev-Popov
ghosts and the temporal and longitudinal components of the gauge fields.
This compensation can be proved diagrammatically \cite{thooft} by means of
the Ward identities or more formally at the level of the Fock space \cite%
{brsunitarity}. For a variety of reasons, we believe that the last word has
not been said on this topic and that an attempt to reorganize and generalize
the proof is most welcome. First, a treatment of perturbative unitarity in
quantum gravity is still missing. Second, the existing proofs in gauge
theories are involved, which suggests that they are not optimized.

In this paper we offer a more economic approach and a new, exhaustive proof
that works not only in Abelian and non-Abelian gauge theories, but also in
quantum gravity and a variety of other nonrenormalizable local theories, in
arbitrary dimensions $d$ greater than 3.

First we prove a number of basic tools, such as the largest time equation
and the cutting equations, paying attention to the minimum assumptions that
they require. Then, we show that the unphysical degrees of freedom can be
consistently dropped from the cuts. To achieve this goal, we identify a 
\textit{special gauge} that leads to the unitarity equation in a
straightforward way.

In gauge theories and gravity, several common gauges have inconvenient
features. The Lorenz gauge, for example, gives propagators that have double
poles, which prevents the derivation of the cutting equations. The Coulomb
gauge has a nice feature, because it just propagates the physical degrees of
freedom. However, it introduces unwanted singularities in the Feynman
diagrams, which are under control only in QED.

The special gauge is a new gauge that has no double poles, interpolates
between the Feynman gauge and the Coulomb gauge and satisfies all the
assumptions that are required to derive the cutting equations. Moreover,
when the gauge fields are given a mass to regulate their on shell infrared
divergences and the Coulomb limit is approached, the threshold for the
production of unphysical particles grows enough to drop them\ out from the
cuts. After that, a few technical tricks allow us to complete the proof. The
special gauge is unique in both Yang-Mills theory and gravity.

An even simpler proof of perturbative unitarity is available in QED, where
it is possible to work directly in the Coulomb gauge.

We pay attention to details such as the regularization and the
renormalization of the cutting equations, the presence of contact terms, the
double poles of the gauge field propagators, the orders of the limits with
which various parameters are removed, the infrared divergences and other
singularities that disappear by summing up the cut diagrams. Some of these
problems are not treated carefully (or are not even mentioned) in the
existing literature.

Because of the key role played by the largest time equation, the proofs we
give in this paper do not generalize to nonlocal quantum field theories of
gauge fields and gravity, including those whose propagators have no poles on
the complex plane besides the graviton one \cite{kuzmin}. For this reason,
the consistency of those theories remains unclear.

The paper is organized as follows. In section \ref{s1} we derive the cutting
equations under the minimum assumptions. In section \ref{s2} we derive the
unitarity equation. In section \ref{specialgauge} we introduce the special
gauge in Abelian and non-Abelian gauge theories. In section \ref{qed} we
give the simplest proof of perturbative unitarity, using the Coulomb gauge
in QED. In section \ref{nonabe} we use the special gauge to prove unitarity
in all gauge theories. In section \ref{qg} we generalize the special gauge
and the proof of unitarity to quantum gravity. Section \ref{conclusions}
contains the conclusions.

\section{The cutting equations}

\label{s1}

\setcounter{equation}{0}

We investigate perturbative unitarity following the guidelines of refs. \cite%
{veltman,diagrammar,diagrammatica}, which consist of proving, in the order:

--- the largest time equation,

--- the cutting equations,

--- the pseudounitarity equation,

--- the unitarity equation.

The pseudounitarity equation is a more general version of the unitarity
equation, where the cuts may propagate both physical and unphysical
particles. In gauge theories and gravity it is helpful to first derive the
pseudounitarity equation and then prove that it implies the unitarity
equation, by showing that the external legs and the cuts can be consistently
projected onto the physical subspace.

In this section we reconsider the cutting equations and search for the
minimum assumptions that are necessary to derive them. We assume invariance
under translations and spatial rotations, but we do not assume Lorentz
invariance. Indeed, we need results that can be applied to gauge choices
that violate Lorentz invariance, such as the Coulomb gauge and the special
gauge. We do not assume from the beginning that the theory is local.
Nevertheless, along with the derivation it emerges that the vertices must be
local.

\subsection{Regularization}

\label{regula}

We use the dimensional regularization, or one of its variants \cite%
{chiraldimreg}, directly in Minkowski spacetime. Several assumptions and
arguments of our derivations make no sense unless there are exactly one time
component and one energy component, so we dimensionally continue the space
coordinates, but do not continue the time coordinate.

Let $d$ denote the physical spacetime dimension and $d-\varepsilon $ the
continued one, where $\varepsilon $ is a complex number. Split the continued
spacetime $\mathbb{R}^{d-\varepsilon }$ into the product of the time line $%
\mathbb{R}$ times the continued space $\mathbb{R}^{d-1-\varepsilon }$.
Denote the metric of flat spacetime with $\eta ^{\mu \nu }=$diag$%
(1,-1,-1,\cdots ,-1)$.

Typically, we work with theories whose vertices are local and whose gauge
fixed propagators $\tilde{f}(E,\mathbf{p})$ are equal to ratios of
polynomials $u(E,\mathbf{p})$ and $v(E,\mathbf{p})$ of the energy $E$ and
the space momentum $\mathbf{p}$, 
\begin{equation}
\tilde{f}(E,\mathbf{p})=\frac{u(E,\mathbf{p})}{v(E,\mathbf{p})},
\label{prostru}
\end{equation}%
with denominators $v(E,\mathbf{p})$ equal to products of polynomials $%
aE^{2}-b\mathbf{p}^{2}-m^{2}+i\epsilon $, where $a$ and $b$ are positive
constants and $m$ is real. The symbol $\epsilon $ is used to specify the
contour prescription.

These theories are well regularized by the prescription of first integrating
on the space momenta $\mathbf{p}$, then on the energies $E$. Indeed, after
the integration on the space momenta the energy integrals behave as 
\begin{equation}
\sim \int^{E=\pm \infty }\mathrm{d}E\hspace{0.01in}\frac{E^{m}}{%
(E^{2})^{n\varepsilon /2}}  \label{beha}
\end{equation}%
for large $|E|$, where $n$ and $m$ are nonnegative integers and $n\neq 0$.
The analytic continuation in $\varepsilon $ makes these integrals well
defined.

Various manipulations can simplify the propagators and generate local
integrands. Then the result is zero, because the dimensionally regularized $%
\mathbf{p}$ integral vanishes, as in%
\begin{equation}
\int_{-\infty }^{+\infty }\frac{\mathrm{d}E}{2\pi }\int \frac{\mathrm{d}%
^{d-1-\varepsilon }\mathbf{p}}{(2\pi )^{d-1-\varepsilon }}%
E^{m}p_{i_{1}}\cdots p_{i_{n}}=0,  \label{deltaf}
\end{equation}%
where, again, $n$ and $m$ are nonnegative integers.

Note that we may not be able to perform the usual contour integrations on
the energy. Nonetheless, each step of the calculation is consistent. For
example, in $d=4$, we have 
\begin{eqnarray}
\int_{-\infty }^{+\infty }\frac{\mathrm{d}E}{2\pi }\int \frac{\mathrm{d}%
^{3-\varepsilon }\mathbf{p}}{(2\pi )^{3-\varepsilon }}\frac{i}{E^{2}-\mathbf{%
p}^{2}-m^{2}+i\epsilon } &=&-\frac{i\Gamma \left( \frac{\varepsilon -1}{2}%
\right) }{(4\pi )^{(3-\varepsilon )/2}}\int_{-\infty }^{+\infty }\frac{%
\mathrm{d}E}{2\pi }(m^{2}-E^{2}-i\epsilon )^{(1-\varepsilon )/2}  \notag \\
&=&\frac{\Gamma \left( \frac{\varepsilon }{2}-1\right) }{(4\pi
)^{(2-\varepsilon )/2}}(m^{2}-i\epsilon )^{1-\frac{\varepsilon }{2}}.
\label{2ex}
\end{eqnarray}

Interchanging the energy and momentum integrals does not make sense, in
general, as in (\ref{deltaf}), but in specific cases it may be allowed, as
in (\ref{2ex}).

The propagators have the structure (\ref{prostru}) in all the cases we
consider, with two exceptions: the Coulomb gauge in QED and the mass terms
introduced to regulate the (on shell) infrared divergences in non-Abelian
gauge theories and nonrenormalizable theories. In both cases some
denominators have $a=0$, but the regularization can be proved to work well
by means of \textit{ad hoc} methods and/or appropriate truncations.

Equipped with this regularization technique, we are ready to begin our
investigation. The algorithm to renormalize the divergences is described
along the way.

\subsection{The largest time equation}

The largest time equation is implied by the following minimum assumptions:

($a$) the vertices are localized in time;

($b$) the propagators $f(x)$ in coordinate space can be decomposed as%
\begin{equation}
f(x)=\theta (x^{0})g_{+}(x)+\theta (-x^{0})g_{-}(x)  \label{teta}
\end{equation}%
in the sense of distributions.

For the moment, we do not make further assumptions about the distributions $%
g_{\pm }(x)$. Formula (\ref{teta}) and similar formulas written below are
exact identities among distributions. In particular, they imply that $f(x)$
contains no contributions proportional to $\delta (x^{0})$ or its
derivatives.

By assumption ($a$), each vertex is associated with a definite time $x^{0}$,
but need not be associated with a unique space coordinate $\mathbf{x}$. By
translational invariance, a propagator is described by a time difference $%
x^{0}-y^{0}$ and a space difference $\mathbf{x}-\mathbf{y}$, as usual.

Consider a raw Feynman diagram in coordinate space. By this we mean the
plain product of the vertices and the propagators, with no integrations over
the space and time coordinates. We denote the raw diagram by $%
F(x_{1}^{0},\cdots ,x_{n}^{0})$, where $x_{i}^{0}$ are the locations of the
vertices in time, while the dependences on the space coordinates are omitted.

Next, build variants $F_{M}$ of the diagram $F$ as follows. Mark any subset
of vertices by putting hats on their times $x_{i}^{0}$. Multiply by an
overall factor $(-1)^{s}$, where $s$ is the number of marked vertices.
Replace the propagators connecting two unmarked vertices, two marked
vertices and a marked vertex with an unmarked one, respectively, as
specified by the following scheme: 
\begin{eqnarray}
x &\longrightarrow &y:\quad \theta (x^{0}-y^{0})g_{+}(x-y)+\theta
(y^{0}-x^{0})g_{-}(x-y),  \notag \\
\hat{x} &\longrightarrow &\hat{y}:\quad \theta
(x^{0}-y^{0})g_{-}(x-y)+\theta (y^{0}-x^{0})g_{+}(x-y),  \label{gator} \\
\hat{x} &\longrightarrow &y:\quad g_{+}(x-y),\qquad \qquad \quad
x\longrightarrow \hat{y}:\quad g_{-}(x-y).  \notag
\end{eqnarray}%
Finally, do not modify the values of the vertices. For the sake of
generality we assume that the propagators are oriented. The orientation is
specified by the arrows.

Now, assume that the vertices have distinct times. Then, we have the identity%
\begin{equation}
\sum_{\text{markings }M}F_{M}(x_{1}^{0},\cdots ,\hat{x}_{i}^{0},\cdots ,\hat{%
x}_{j}^{0},\cdots ,x_{n}^{0})=0,  \label{lte}
\end{equation}%
which is known as the largest time equation. The sum is over all the ways to
mark the vertices, including the cases where the vertices are all marked and
all unmarked.

Here is the proof of (\ref{lte}). Since the vertices have distinct times,
one vertex must have the largest time. Denote that time by $z^{0}$. Pick any
diagram $F_{\bar{M}}$ of the sum (\ref{lte}). The time $z^{0}$ may be marked
or not in $F_{\bar{M}}$. If it is marked (unmarked), the sum (\ref{lte})
contains another diagram $F_{\bar{M}}^{\prime }$ that is identical to $F_{%
\bar{M}}$ except for the fact that $z^{0}$ is unmarked (marked). The sum $F_{%
\bar{M}}+F_{\bar{M}}^{\prime }$ vanishes, because the propagators between a
point\footnote{%
There maybe more than one point with time $z^{0}$, if the vertex is nonlocal
in space.} $z=(z^{0},\mathbf{z})$ and any other points $x$, $y$ are, in the
various cases, 
\begin{eqnarray*}
z &\longrightarrow &y:\ g_{+}(z-y),\qquad \hat{z}\longrightarrow y:\
g_{+}(z-y),\qquad z\longrightarrow \hat{y}:\ g_{-}(z-y),\qquad \hat{z}%
\longrightarrow \hat{y}:\ g_{-}(z-y), \\
x &\longrightarrow &z:\ g_{-}(x-z),\qquad x\longrightarrow \hat{z}:\
g_{-}(x-z),\qquad \hat{x}\longrightarrow z:\ g_{+}(x-z),\qquad \hat{x}%
\longrightarrow \hat{z}:\ g_{+}(x-z).
\end{eqnarray*}%
In the end, the diagrams $F_{\bar{M}}$ and $F_{\bar{M}}^{\prime }$ are equal
except for an overall minus sign due to the marking/unmarking of $z^{0}$.
This implies (\ref{lte}).

\subsection{Contact terms}

\label{contact}

To derive the cutting equations, we must calculate the Fourier transforms of
the largest time equations, which demands to integrate on the coordinates.
However, in the derivation of (\ref{lte}) we have assumed that the vertices
had different times. We want to make sure that this assumption can be
dropped, because only in that case the result of the Fourier transform has a
straightforward diagrammatic interpretation.

More precisely, we need to show that when we take any (one-sided) limits of
coinciding times on the functions $F_{M}$ of equation (\ref{lte}), we do not
miss terms that give nontrivial contributions to the integrals on the
coordinates.

Call two vertices nearest neighbors if they are connected by a propagator.
Observe that, to prove (\ref{lte}), the point $z$ of largest time just needs
to be compared with its nearest neighbors. For this reason, equation (\ref%
{lte}) trivially extends to the case where there are vertices with
coinciding times, as long as no pairs of them are made of nearest neighbors.
Precisely, denote the vertices with coinciding times by $w_{i}$ and call
their time $w^{0}$. When $w^{0}$ is not the largest time, we can proceed
exactly as above, which leads to (\ref{lte}). When $w^{0}$ is the largest
time, we can pick any of the $w_{i}$ as the vertex $z$ and, again, proceed
as above to obtain (\ref{lte}). Thus, the only situation that deserves
attention is when some nearest neighbors have coinciding times. Nontrivial
contributions to the integrals on the coordinates can only appear when
contact terms are present.

Consider that the vertices may carry time derivatives. For example, in
quantum gravity the Einstein-Hilbert action is corrected by terms built with
the Riemann tensor and its derivatives, which may contain an arbitrary
number of time derivatives acting on the metric tensor. By means of partial
integrations, the derivatives can be moved to the propagators (\ref{gator}).
Then, they may generate contact terms proportional to $\delta (x^{0})$ or
its derivatives: 
\begin{equation}
\partial _{0}^{n}f(x)=\theta (x^{0})\partial _{0}^{n}g_{+}(x)+\theta
(-x^{0})\partial _{0}^{n}g_{-}(x)+\sum_{k=1}^{n}\delta ^{(k-1)}(x^{0})\left[
\lim_{x^{0}\rightarrow 0^{+}}\partial
_{0}^{n-k}g_{+}(x)-\lim_{x^{0}\rightarrow 0^{-}}\partial _{0}^{n-k}g_{-}(x)%
\right] .  \label{conta}
\end{equation}%
However, the largest time equation (\ref{lte}) is only sensitive to the
first two terms that appear on the right-hand side of this equation, since
the vertices must have distinct times.

In specific cases, such as when the vertices cannot provide enough time
derivatives to create nontrivial contact terms, assumption ($a$) is
sufficient for our purposes. However, in general it is necessary to replace
it with the stronger assumption that

($a^{\prime }$) the vertices are local

\noindent and further assume that

($c$) the contact terms are local, i.e. \noindent the time derivatives of
the propagators satisfy the property 
\begin{equation}
\partial _{0}^{n}f(x)=\theta (x^{0})\partial _{0}^{n}g_{+}(x)+\theta
(-x^{0})\partial _{0}^{n}g_{-}(x)+\text{local terms};  \label{inut}
\end{equation}

($d$) $g_{\pm }(x)$ and their derivatives $\partial _{0}^{n}g_{\pm }(x)$
have well-defined limits for $x^{0}\rightarrow 0$.

When the propagators have the structure (\ref{prostru}) property ($c$)
follows as a consequence. Observe that a nontrivial contact term arises when
the numerator contains a power of $E$ greater than or equal to the maximum
power of $E$ appearing in the denominator. Let $r$ denote the degree of $v(E,%
\mathbf{p})$ as a polynomial in $E$. Assumption ($b$) implies that the
degree of $u(E,\mathbf{p})$ in $E$ must be smaller than $r$. Write 
\begin{equation*}
\tilde{f}(E,\mathbf{p})=\frac{u(E,\mathbf{p})}{E^{r}+w(E,\mathbf{p})},
\end{equation*}%
where $w(E,\mathbf{p})$ also has degree smaller than $r$. When $\tilde{f}(E,%
\mathbf{p})$ is multiplied by a sufficient power of $E$, the numerator may
contain a power $E^{r}$ that simplifies the power $E^{r}$ of the
denominator. Write $u(E,\mathbf{p})=E^{r-1}u^{\prime }(\mathbf{p})+u^{\prime
\prime }(E,\mathbf{p})$, where $u^{\prime }(\mathbf{p})$ is a polynomial of $%
\mathbf{p}$ and $u^{\prime \prime }(E,\mathbf{p})$ is a polynomial of degree 
$r-2$ in $E$. Then,%
\begin{equation*}
E\tilde{f}(E,\mathbf{p})=\frac{Eu^{\prime \prime }(E,\mathbf{p})-w(E,\mathbf{%
p})u^{\prime }(\mathbf{p})}{E^{r}+w(E,\mathbf{p})}+u^{\prime }(\mathbf{p}).
\end{equation*}%
The ratio on the right-hand side does not contain contact terms, because the
numerator contains at most $r-1$ powers of $E$. Thus, (\ref{inut}) holds for 
$n=1$. The argument can be easily iterated for $E^{n}\tilde{f}(E,\mathbf{p})$%
, $n>1$, which proves (\ref{inut}) for every $n$.

It is easy to show that property ($d$)\ follows from (\ref{prostru}), as
long as (\ref{prostru}) has only simple poles and $g_{\pm }$ are regular
distributions.

We are ready to describe the procedure to deal with the contact terms.
Consider a diagram $G$ where some differentiated propagators carry contact
terms. Separate them from the rest of $\partial _{0}^{n}f(x)$ as in formula (%
\ref{inut}) and write $G$ as a sum of contributions $G_{\text{c}}$, such
that each internal line of $G_{\text{c}}$ is either a contact term or does
not carry contact terms. The contact terms of $G_{\text{c}}$ draw a
subdiagram $G_{\text{sub}}$ (which may be disconnected), as shown in the
picture%
\begin{equation*}
\includegraphics[width=10truecm]{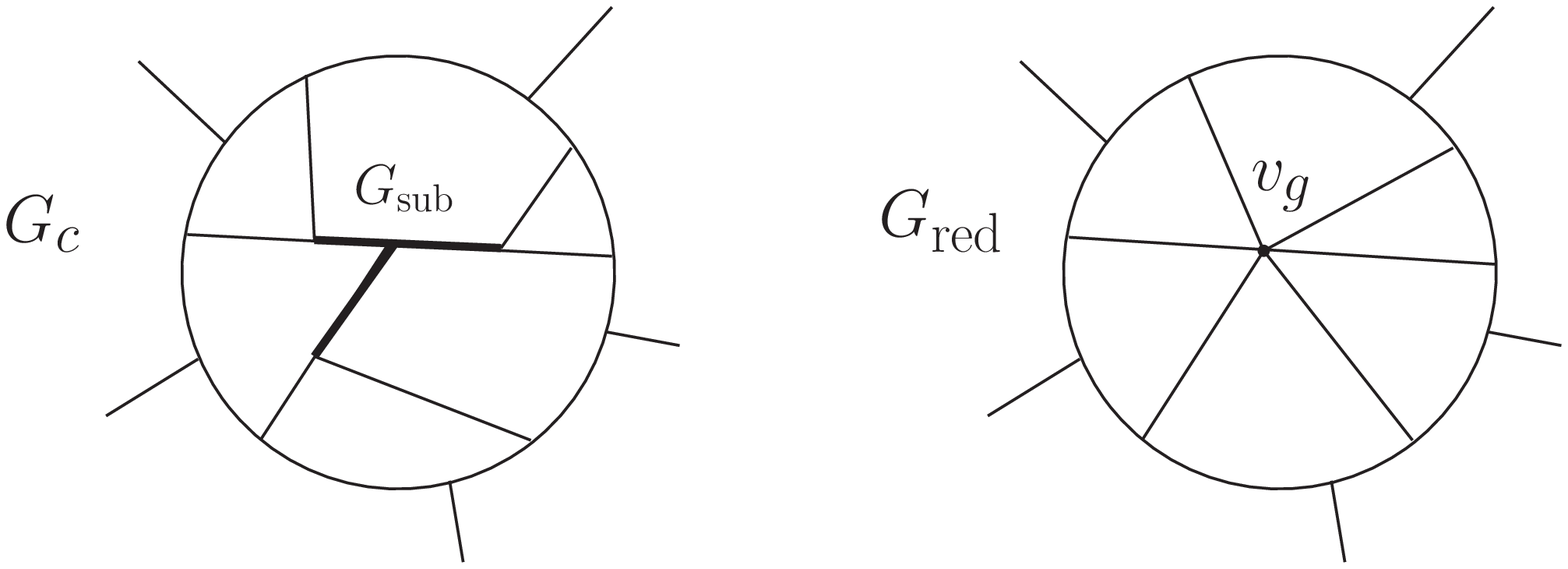}
\end{equation*}
It is easy to prove that if $G_{\text{sub}}$ contains loops, it vanishes.
Indeed, by assumptions ($a^{\prime }$) and ($c$), the vertices and the
contact terms are both local, so each loop of contact terms is a linear
combination of integrals (\ref{deltaf}) in momentum space. Thus, we can
assume that $G_{\text{sub}}$ is a tree subdiagram. Each connected component $%
G_{\text{sub}}^{\text{conn}}$ of $G_{\text{sub}}$ is equal to a product of
(derivatives of) delta functions $\delta (x_{i}^{0}-x_{j}^{0})$ times a new
local vertex $v_{\text{g}}$ that can be obtained by gluing the vertices of $%
G_{\text{sub}}^{\text{conn}}$ together. In turn, $G_{\text{c}}$ is a product
of (derivatives of) delta functions $\delta (x_{i}^{0}-x_{j}^{0})$ times a
reduced diagram $G_{\text{red}}$, built with the ordinary vertices and the
vertices $v_{\text{g}}$.

Now, $G_{\text{red}}$ has no contact terms and thus satisfies the largest
time equation (\ref{lte}). It is connected if the original diagram $G$ is
connected.

By property ($d$), a line that connects a marked vertex with an unmarked one
is not interested by contact terms. By the same property, $\partial
_{0}^{n}g_{\pm }(x)$ have well-defined limits for $x^{0}\rightarrow 0$.
Then, formula (\ref{gator}) shows that the contact terms carried by the
lines connecting pairs of marked vertices are equal to minus the contact
terms of $\partial _{0}^{n}f(x)$. We can easily show that, thanks to this
fact, a minus sign is associated with each marked vertex of type $\hat{v}_{%
\text{g}}$, as expected. Indeed, $\hat{v}_{\text{g}}$ originates from the
markings of all the vertices of $G_{\text{sub}}^{\text{conn}}$. Each such
vertex provides a minus sign, but other minus signs come from the contact
terms of $G_{\text{sub}}^{\text{conn}}$, because they are associated with
pairs of marked vertices. Since $G_{\text{sub}}^{\text{conn}}$ is a tree
diagram, the sum of the number of its vertices plus the number of its lines
is odd, so $\hat{v}_{\text{g}}$ always carries a minus sign.

If we sum the $G$ largest time equation (\ref{lte}), derived under the
condition that all the nearest neighbors have distinct times, to the largest
time equations satisfied by the diagrams $G_{\text{red}}$, multiplied by the
appropriate products of (derivatives of) delta functions $\delta
(x_{i}^{0}-x_{j}^{0})$, the right-hand side of (\ref{conta}) is fully
reconstructed, for each propagator. Observe that assumption ($a$) plays an
important role here, because it ensures that each diagram involves a finite
number of time derivatives.

The conclusion is that if we add assumptions ($a^{\prime }$), ($c$) and ($d$%
), the largest time equation (\ref{lte}) holds even if we drop the
assumption that the vertices are located at distinct times. Then formula (%
\ref{lte}) can be interpreted as an identity of distributions and we can
safely compute its Fourier transform.

The same conclusion holds when the vertices do not provide enough time
derivatives to generate contact terms, in which case assumptions ($a^{\prime
}$), ($c$) and ($d$) need not be satisfied.

\subsection{The cutting equations}

Once the contact terms are dealt with as explained above, the Fourier
transform of the largest time equation (\ref{lte}) is an analogous equation
in momentum space, where the propagators and the vertices are replaced by
their Fourier transforms. Denoting the Fourier transform of $F_{M}$ with $%
G_{M}$, we get%
\begin{equation}
\sum_{\text{markings }M}G_{M}(p_{1},\cdots ,p_{n})=0.  \label{ltef0}
\end{equation}

Now we simplify this identity by converting it into a set of cutting
equations. The cutting equations are consequences of the assumptions made so
far and the following additional one:

($e$) the Fourier transforms $\tilde{g}_{\pm }(p)$ of $g_{\pm }(x)$ have the
form%
\begin{equation}
\tilde{g}_{\pm }(p)=\theta (\pm p^{0})h_{\pm }(p).  \label{assd}
\end{equation}

For the moment, we make no further assumptions about the distributions $%
h_{\pm }(p)$. We interpret formulas (\ref{assd}) by saying that the energy
flows from an unmarked vertex to a marked vertex, that is to say from\ the
past to the future.

Consider a connected, amputated diagram of formula (\ref{ltef0}). Call the
external legs whose energies flow into (out of) the diagram ingoing
(outgoing). Mark the end points of the outgoing external legs and leave the
end points of the incoming external legs unmarked.

We refer to the vertices and the end points of the external legs by simply
calling them \textquotedblleft points\textquotedblright . Thus, the energy
flows from an unmarked point to a marked point. Between two marked points or
two unmarked points it can flow in both directions.

We want to show that every diagram $G_{M}$ of (\ref{ltef0}) vanishes, unless
it can be cut into two pieces, leaving the marked and unmarked points on
opposite sides of the cut. If that is the case, we denote the diagram by $%
G_{C}$.

Consider a marked vertex. Its nearest points cannot be all unmarked, because
then the orientations of the energy flows would imply the violation of
energy conservation. Thus, at least one of its nearest neighbors is a marked
point. Next, consider a connected subdiagram made of some marked vertices
and the legs attached to them. Again, energy conservation implies that the
nearest points of the subdiagram must include at least another marked point.
Extending the subdiagram point by point, we find that each connected
subdiagram of marked points must include the end point of an outgoing line.
Similarly, a connected subdiagram of unmarked points must include the end
point of an incoming line.

Because of this, the diagram is cut into two (not necessarily connected)
subdiagrams. The cut crosses the propagators that connect a marked vertex to
an unmarked vertex, as well as the external lines that connect a marked
point to an unmarked point. For example, we have%
\begin{equation}
\includegraphics[width=10truecm]{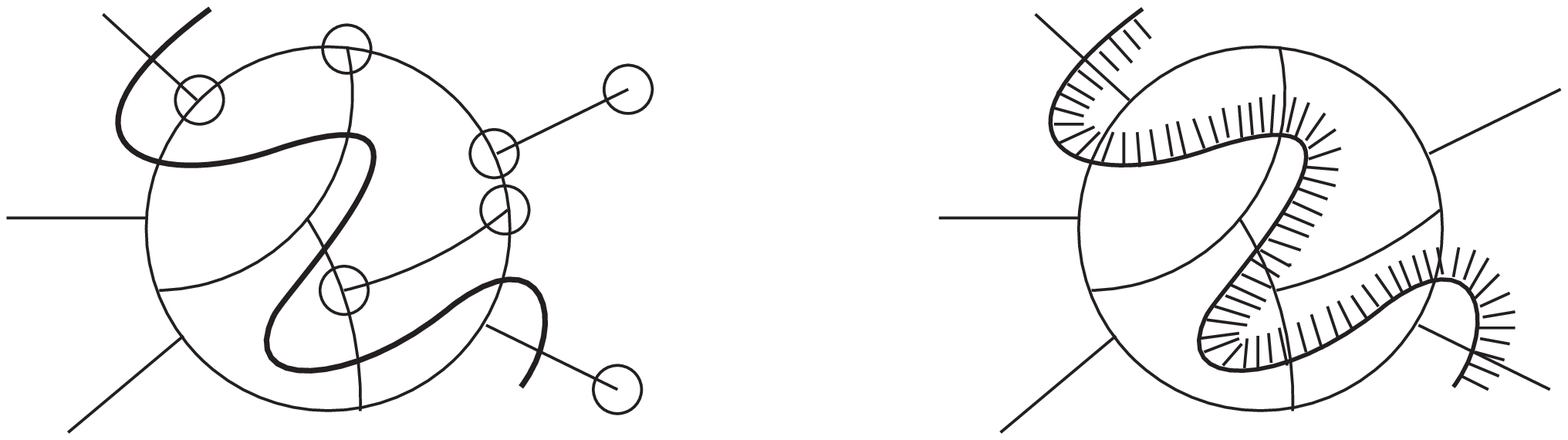}  \label{cut}
\end{equation}%
In the left figure, the marked points are circled and the solid line denotes
the cut. From now on, instead of marking the vertices, we just shadow the
marked side of the cut, as shown in the right figure of (\ref{cut}).
Normally, the incoming legs are drawn on the left-hand side and the outgoing
legs are drawn on the right-hand side.

Since the external legs are amputated, the cutting of an external leg does
not have any particular meaning besides the graphical one: all the marked
points must lie on one side of the cut and all the unmarked points must lie
on the other side.

We conclude that the Fourier transform (\ref{ltef0}) of the largest time
equation (\ref{lte}) simplifies into the cutting equation%
\begin{equation}
\sum_{\text{cuttings }C}G_{C}(p_{1},\cdots ,p_{n})=0.  \label{ltef}
\end{equation}%
We stress that equations (\ref{lte}), (\ref{ltef0}) and (\ref{ltef}) do not
assume that the external legs are on shell.

The sum of formula (\ref{ltef}) contains two special contributions that it
is convenient to single out. They are the contributions $G$ and $\bar{G}$ of
the diagrams where all the vertices are unmarked or marked, respectively. We
have%
\begin{equation}
G(p_{1},\cdots ,p_{n})+\bar{G}(p_{1},\cdots ,p_{n})=-\sum_{\text{proper
cuttings }C}G_{C}(p_{1},\cdots ,p_{n}),  \label{ltefr}
\end{equation}%
where the sum is restricted to the \textquotedblleft
proper\textquotedblright\ cuttings, which are those where at least one
vertex is marked and at least one vertex is unmarked.

Everything we have said so far is valid at the regularized level. If the
locality of counterterms holds, the diagrams built with the counterterms
satisfy analogous properties. Combining the cutting equation of one diagram
with the cutting equations satisfied by the diagrams that subtract its
subdivergences and overall divergence, we obtain the renormalized cutting
equation.

Note that in the renormalized cutting equation every side of the cut is
appropriately renormalized. On the other hand, no counterterms are
associated with subdiagrams containing the cut or part of it. The
consistency of this fact is proved by the renormalized cutting equation
itself. Indeed, after the inclusion of the counterterms the left-hand side
of formula (\ref{ltefr}) is convergent, so the right-hand side must also be
convergent.

So far, the assumptions we have made are more general than the usual ones.
However, we anticipate that we cannot obtain the pseudounitarity equation
unless we impose further restrictions.

\subsection{Examples}

Now we give some simple examples concentrating on scalar fields $\varphi $.
Examples with fermions and gauge fields are given later on.

If we interpret the decomposition (\ref{teta}) as the usual T-ordered one,
where%
\begin{equation*}
f(x)=\langle 0|T\varphi (x)\varphi (0)|0\rangle ,\qquad g_{+}(x)=\langle
0|\varphi (x)\varphi (0)|0\rangle ,\qquad g_{-}(x)=\langle 0|\varphi
(0)\varphi (x)|0\rangle =g_{+}(-x),
\end{equation*}%
and further assume Lorentz invariance, then we obtain the standard K\"{a}ll%
\'{e}n-Lehman (KL) representation. Indeed, now $f(x)$ and $g_{\pm }(x)$ are
Lorentz invariant and so are $\tilde{g}_{\pm }(p)=\theta (\pm p^{0})h_{\pm
}(p)$. However, the sign of $p^{0}$ depends on the reference frame, unless $%
p^{2}>0$. Thus, $h_{\pm }(p)$ vanish for $p^{2}<0$ and depend only on $p^{2}$
for $p^{2}>0$. Then, $g_{-}(x)=g_{+}(-x)$ implies that $h_{+}$ and $h_{-}$
must be the same function, which we denote by $(2\pi )\rho (p^{2})$.
Inserting $\tilde{g}_{\pm }(p)=(2\pi )\theta (\pm p^{0})\rho (p^{2})$ inside
(\ref{teta}) and working out the Fourier transform $\tilde{f}(p)$ of $f(x)$,
we find the KL decomposition 
\begin{equation}
\tilde{f}(p)=\int_{0}^{+\infty }\frac{i\rho (s)\mathrm{d}s}{%
p^{2}-s+i\epsilon }.  \label{commonKL}
\end{equation}%
We have used the identity%
\begin{equation}
\theta (x^{0})=\frac{i}{2\pi }\int_{-\infty }^{+\infty }\frac{\mathrm{e}%
^{-i\tau x^{0}}\mathrm{d}\tau }{\tau +i\epsilon },  \label{tt}
\end{equation}%
then changed the integration variable from $\tau $ to $\tau ^{2}-\mathbf{p}%
^{2}$ and used $\rho (p^{2})=0$ for $p^{2}<0$. Note that $\rho $ is not
assumed to be nonnegative.

In the case of ordinary (i.e. non-higher-derivative) free scalar fields, we
have%
\begin{equation*}
\tilde{f}(p)=\frac{i}{p^{2}-m^{2}+i\epsilon },\qquad \tilde{g}_{\pm
}(p)=2\pi \theta (\pm p^{0})\rho (p^{2}),\qquad \rho (s)=\delta (s-m^{2}).
\end{equation*}%
The simplest cutting equation is the one satisfied by the propagator:%
\begin{equation}
\includegraphics[width=12truecm]{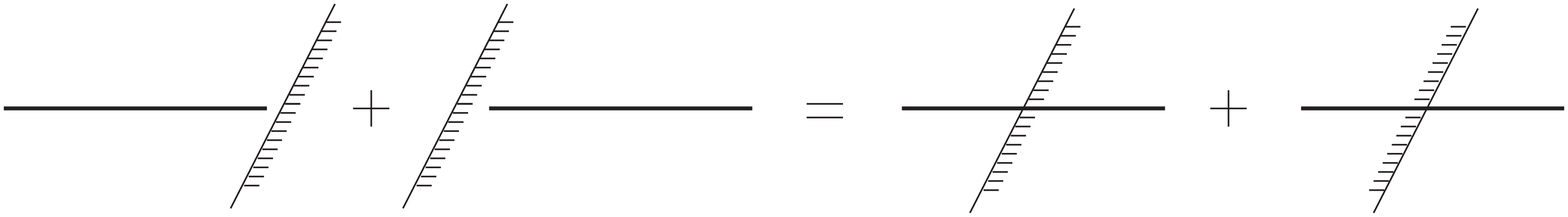}  \label{c1}
\end{equation}%
In deriving this equation, the end points must be imagined as vertices, so
each shadowed end point gives a factor $-1$. This explains the signs of (\ref%
{c1}). In formulas, we have%
\begin{equation}
\frac{i\hspace{0.01in}}{p^{2}-m^{2}+i\epsilon }+\frac{-i\hspace{0.01in}}{%
p^{2}-m^{2}-i\epsilon }=2\pi \theta (p^{0})\delta (p^{2}-m^{2})+2\pi \theta
(-p^{0})\delta (p^{2}-m^{2}).  \label{c1f}
\end{equation}

At one loop we have%
\begin{equation}
\includegraphics[width=14truecm]{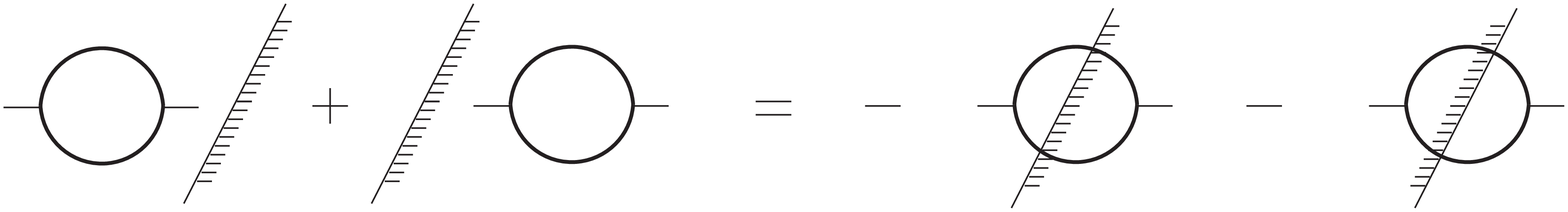}  \label{c2}
\end{equation}%
which can be checked easily (see for example \cite{diagrammatica}).

Nonlocal quantum field theories do not satisfy the assumptions that lead to
the largest time equation, unless their vertices are local in time. Then,
however, either Lorentz invariance or gauge invariance is violated.

\section{The pseudounitarity and unitarity equations}

\label{s2}

\setcounter{equation}{0}

In this section we derive the pseudounitarity equation and explain when it
implies the unitarity equation. As said, we must make additional
assumptions, which eventually lead to a general K\"{a}ll\'{e}n-Lehman
spectral representation, even if we do not assume it from the start.

First, the shadowed regions of the cutting equations should correspond to
the complex conjugate diagrams, that is to say we must assume that

($f$) the action is Hermitian.

In particular, this implies that the shadowed propagator is the Hermitian
conjugate of the unshadowed one and that the cut propagators (\ref{assd})
are Hermitian, i.e. $\tilde{g}_{\pm }^{\dagger }(p)=\tilde{g}_{\pm }(p)$.
The minus sign associated with each marked vertex is then justified by the
fact that the vertices are anti-Hermitian.

Second, the cut propagators must project onto the on shell states of the
free field limit. This means that we must replace ($e$) by the more
restrictive assumption that

($e^{\prime }$) the Fourier transforms $\tilde{g}_{\pm }(p)$ of $g_{\pm }(x)$
have the form 
\begin{equation}
\tilde{g}_{\pm }(p)=\pi \sum_{i=1}^{N}a_{i}^{\pm }(p)\delta (p^{0}\mp \omega
_{i}),  \label{gp}
\end{equation}%
where $\omega _{i}(\mathbf{p}^{2}\mathbf{)}$ are positive functions of $%
\mathbf{p}$, while $a_{i}^{\pm }(p)$ are Hermitian matrices whose entries
are functions of $p$.

Using the identity (\ref{tt}) it is easy to check that the Fourier transform
of formula (\ref{teta}) is%
\begin{equation}
\tilde{f}(p)=\frac{i}{2}\sum_{i=1}^{N}\frac{a_{i}^{+}(\omega _{i},\mathbf{p}%
)(p^{0}+\omega _{i})-a_{i}^{-}(-\omega _{i},\mathbf{p})(p^{0}-\omega _{i})}{%
(p^{0})^{2}-\omega _{i}^{2}+i\epsilon }.  \label{propa}
\end{equation}%
With suitable assumptions on $a_{i}^{\pm }$ and $\omega _{i}$, this formula
matches (\ref{prostru}).

At this point, we diagonalize the matrices $a_{i}^{+}(\omega _{i},\mathbf{p}%
) $ and $a_{i}^{-}(-\omega _{i},\mathbf{p})$ and normalize their eigenvalues
to $1$, $-1$ and $0$. Calling the diagonalizing matrices $u_{i}(\mathbf{p)}$
and $v_{i}(\mathbf{p)}$, we have%
\begin{equation*}
a_{i}^{+}(\omega _{i},\mathbf{p})=u_{i}(\mathbf{p)}H_{i}u_{i}^{\dagger }(%
\mathbf{p}),\qquad a_{i}^{-}(-\omega _{i},-\mathbf{p})=(-1)^{\sigma
_{i}}v_{i}(\mathbf{p)}H_{i}^{\prime }v_{i}^{\dagger }(\mathbf{p}),
\end{equation*}%
where $\sigma _{i}=0$, $1$ for bosons and fermions, respectively, and $H_{i}$%
, $H_{i}^{\prime }$ are diagonal matrices with eigenvalues $1$, $0$ and $-1$%
. The matrices $u_{i}(\mathbf{p)}$ and $v_{i}(\mathbf{p)}$ collect the
external particle and antiparticle states.

Writing the $S$ matrix as $S=1+iT$, the cutting equations (\ref{ltefr}) can
be collected into the pseudounitarity equation 
\begin{equation}
-iT+iT^{\dagger }=THT^{\dagger },  \label{puni}
\end{equation}%
where $H$ is the diagonal matrix having diagonal blocks $H_{i}$ and $%
H_{i}^{\prime }$.

If there exists a subspace $V$ of states of the free field theory such that
equation (\ref{puni}) holds with $H=1$ when the external legs and the cut
legs are projected onto $V$, then the pseudounitarity equation implies
perturbative unitarity, which is expressed by the equation%
\begin{equation}
-iT+iT^{\dagger }=TT^{\dagger }  \label{uni}
\end{equation}%
in $V$.

Summarizing, the assumption that turns the pseudounitarity equation into the
unitarity equation is that

($g$) there exists a subspace $V$ of states with $a_{i}^{+}(\omega _{i},%
\mathbf{p})>0$, $(-1)^{\sigma _{i}}a_{i}^{-}(-\omega _{i},-\mathbf{p})>0$,
such that the cutting equations still hold after the external legs and the
cut legs are projected onto $V$.

\subsection{The K\"{a}ll\'{e}n-Lehman spectral representation}

We have found that, in general, the propagator must have the form (\ref%
{propa}), which means, in particular, that there can only be simple poles on
the real axis, but no double poles and no poles away from the real axis. We
can recast formula (\ref{propa}) in the form of the general K\"{a}ll\'{e}%
n-Lehman representation 
\begin{equation}
\tilde{f}(p)=i\int_{0}^{+\infty }\frac{\rho (s,\mathbf{p})+p^{0}\sigma (s,%
\mathbf{p})}{(p^{0})^{2}-s+i\epsilon }\mathrm{d}s,  \label{genKL}
\end{equation}%
where the densities $\sigma $ and $\rho $ are the Hermitian matrices given by%
\begin{eqnarray*}
\rho (s,\mathbf{p}) &=&\sum_{i=1}^{N}\frac{\omega _{i}}{2}\left(
a_{i}^{+}(\omega _{i},\mathbf{p})+a_{i}^{-}(-\omega _{i},\mathbf{p})\right)
\delta (s-\omega _{i}^{2}), \\
\sigma (s,\mathbf{p}) &=&\sum_{i=1}^{N}\frac{1}{2}\left( a_{i}^{+}(\omega
_{i},\mathbf{p})-a_{i}^{-}(-\omega _{i},\mathbf{p})\right) \delta (s-\omega
_{i}^{2}).
\end{eqnarray*}%
It is easy to check formula (\ref{c1}) in this general case.

The representation (\ref{genKL}) has a form similar to the one known from
Lorentz violating theories \cite{confnolor}. When Lorentz symmetry holds,
the densities $\rho (s,\mathbf{p})$ and $\sigma (s,\mathbf{p})$ vanish for $%
s<$ $\mathbf{p}^{2}$, so after a translation the representation acquires a
more common form, that is to say (\ref{commonKL}) with $\rho (s)$ replaced
by $\rho (s+\mathbf{p}^{2},\mathbf{p})+p^{0}\sigma (s+\mathbf{p}^{2},\mathbf{%
p})$. Lorentz invariance also implies that this sum has the form\ $\mathbf{%
p\cdot }\boldsymbol{\rho }^{\prime }(s)+p^{0}\sigma ^{\prime }(s)+\rho
^{\prime \prime }(s)$ and further relates the functions $\boldsymbol{\rho }%
^{\prime }(s)$ and $\sigma ^{\prime }(s)$.

\subsection{Examples}

Most bosons have $a^{-}(-\omega ,\mathbf{p})=a^{+}(\omega ,\mathbf{p})$, so
the coefficient $\sigma (s,\mathbf{p})$ of $p^{0}$ in the numerator of (\ref%
{genKL}) vanishes. This gives 
\begin{equation}
\tilde{f}(p)=i\frac{\omega a^{+}(\omega ,\mathbf{p})}{(p^{0})^{2}-\omega
^{2}+i\epsilon }.  \label{fey}
\end{equation}
Lorentz invariant scalars have $a^{+}(\omega ,\mathbf{p})=1/\omega $, $%
\omega =\sqrt{\mathbf{p}^{2}+m^{2}}$.

Examples where the coefficient of $p^{0}$ does not vanish are the
Chern-Simons gauge fields and the fermions. In particular, free Dirac
fermions have $a^{+}(p)=a^{-}(p)=(p_{\mu }\gamma ^{\mu }+m)\gamma
^{0}/\omega $, which gives%
\begin{equation*}
\tilde{f}(p)\gamma ^{0}=i\frac{p_{\mu }\gamma ^{\mu }+m}{p^{2}-m^{2}+i%
\epsilon }.
\end{equation*}%
Interacting fermions\ in Lorentz invariant theories have%
\begin{equation*}
\tilde{f}(p)\gamma ^{0}=i\int_{0}^{+\infty }\frac{\rho (s)+p_{\mu }\gamma
^{\mu }\sigma (s)}{p^{2}-s+i\epsilon }\mathrm{d}s.
\end{equation*}

Let us now consider gauge fields. If we choose a covariant gauge the
pseudounitarity equation exists only when the\ propagators have the form%
\begin{equation}
\tilde{f}_{\mu _{1}\cdots \mu _{n},\nu _{1}\cdots \nu _{n}}(p)=i\frac{%
\mathcal{I}_{\mu _{1}\cdots \mu _{n},\nu _{1}\cdots \nu _{n}}}{%
p^{2}+i\epsilon },  \label{zero}
\end{equation}%
where $\mathcal{I}_{\mu _{1}\cdots \mu _{n},\nu _{1}\cdots \nu _{n}}$ is a
constant tensor built with the metric $\eta _{\mu \nu }$. In other words, no
covariant gauges besides the Feynman ones satisfy the assumptions. The
common Lorenz gauge for vector fields, which gives the propagator%
\begin{equation}
-\frac{i}{p^{2}}\left( \eta _{\mu \nu }-(1-\lambda )\frac{p_{\mu }p_{\nu }}{%
p^{2}}\right) ,  \label{nocon}
\end{equation}%
does not lead to the cutting equations (\ref{ltefr}) when the gauge-fixing
parameter $\lambda $ is different from $1$, because of the double pole. It
is possible to deform (\ref{nocon}) by introducing fictitious masses that
split the double pole into simple poles, but it is not easy to study the
limit where the fictitious masses are removed in the cutting equations.

We see that, in the end, the conditions imposed by the very existence of the
cutting equations and the requirement that they lead to the pseudounitarity
and unitarity equations are very restrictive.

The largest time equation (\ref{lte}), the cutting equations (\ref{ltefr}),
the pseudounitarity equation (\ref{puni}) and the unitarity equation (\ref%
{uni}) also hold when the external legs of the diagrams correspond to the
insertions of local composite fields. Indeed, it is easy to check that the
arguments that lead to those equations remain valid. More generally, the
equations still hold when the external legs include both elementary fields
and local composite fields.

\subsection{Infrared divergences and other singularities}

\label{infra}

The uncut diagrams that appear on the left-hand side of the cutting
equations (\ref{ltefr}) are regular off shell. However, the individual
diagrams that appear on the right-hand side have cuts, which are necessarily
on shell. In the presence of massless particles there can be infrared
divergences. For example, consider the sum 
\begin{equation}
\includegraphics[width=12truecm]{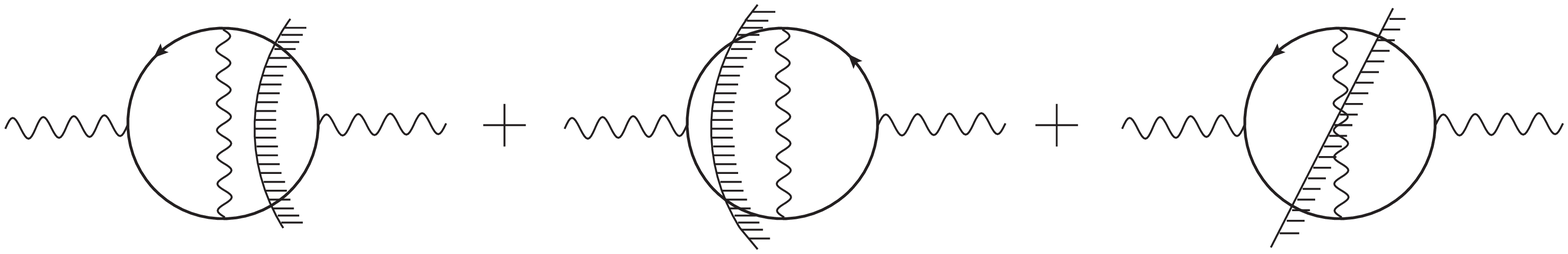}  \label{ir}
\end{equation}%
in QED. The first two diagrams contain the infrared divergences of the
one-loop radiative corrections to the vertex. However, the third diagram is
also infrared divergent and compensates the divergences of the other two.

The cancellation of the infrared divergences on the right-hand side of
equation (\ref{ltefr}) (when the external legs are off shell) is a
well-known fact \cite{infred}, so we do not need to spend more words on it.
At the same time, for various arguments of the next sections we need to deal
with cut diagrams that are individually infrared convergent. This can be
achieved by inserting fictitious masses in the propagators. It is possible
to do so without violating the assumptions we have made so far. However, the
fictitious masses violate gauge invariance and it is necessary to remove
them with care to successfully prove the perturbative unitarity of gauge
theories.

Other singularities occur when self-energy subdiagrams are present. For
example, the product of a cut propagator times an unshadowed propagator with
the same momentum is equal to%
\begin{equation}
\frac{i(2\pi )\theta (p^{0})\delta (p^{2}-m^{2})}{p^{2}-m^{2}+i\epsilon }=%
\frac{2\pi }{\epsilon }\theta (p^{0})\delta (p^{2}-m^{2}),  \label{p1}
\end{equation}%
in the case of ordinary scalar fields. On the other hand, the product of an
unshadowed propagator times a shadowed one with the same momentum is%
\begin{equation}
\frac{i}{p^{2}-m^{2}+i\epsilon }\frac{-i}{p^{2}-m^{2}-i\epsilon }=\frac{1}{%
(p^{2}-m^{2}+i\epsilon )^{2}}+\frac{2\pi }{\epsilon }\delta (p^{2}-m^{2}),
\label{p2}
\end{equation}%
where we have used (\ref{c1f}).

Again, the left-hand sides of the cutting equations are smooth, while the
individual diagrams on the right-hand side may have singularities for $%
\epsilon \rightarrow 0$ that cancel out in the sum. The cancellation can be
seen by keeping the width $\epsilon $ different from zero and taking the
limit $\epsilon \rightarrow 0$ only at the very end.

For example, consider the bubble diagram where one propagator is replaced by
the one-loop self-energy. The right-hand side of the cutting equation is
equal to minus the sum%
\begin{equation}
\includegraphics[width=12truecm]{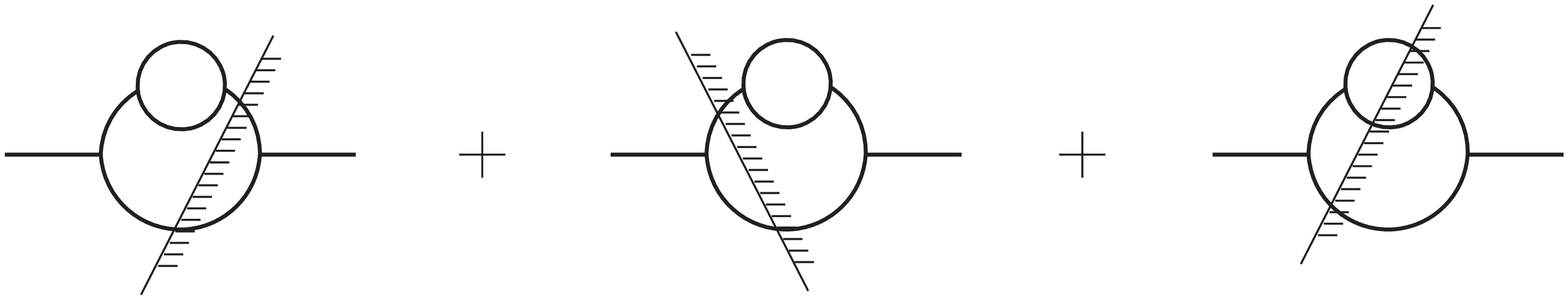}  \label{c4}
\end{equation}%
where we have assumed, for definiteness, that the energy flows in from the
left. Using (\ref{c2}), (\ref{p1}) and (\ref{p2}), it is easy to check that
the sum%
\begin{equation}
\includegraphics[width=12truecm]{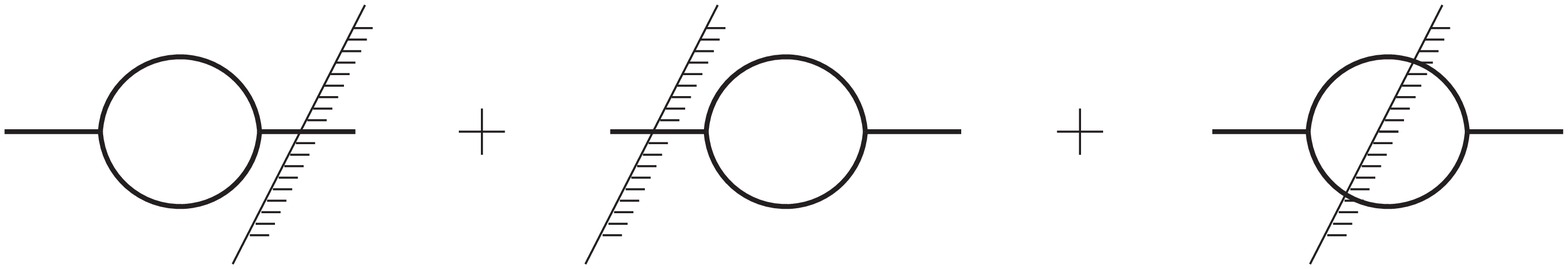}  \label{c3}
\end{equation}%
(with propagators on the external legs) is equal to%
\begin{equation*}
\raisebox{-0.45\height}{\includegraphics[width=2truecm]{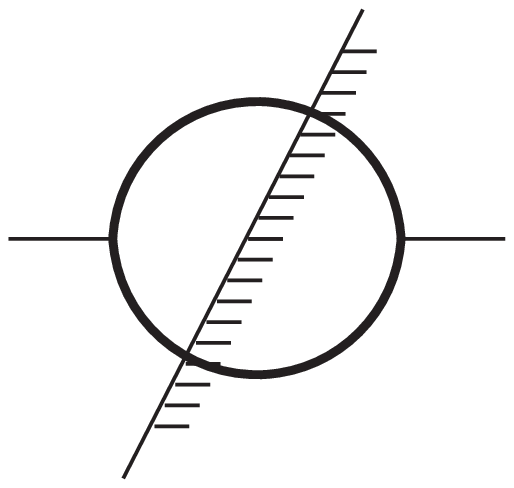}}\times \frac{%
1}{(p^{2}-m^{2}+i\epsilon )^{2}},
\end{equation*}%
which is regular when $\epsilon \rightarrow 0$. Thus, (\ref{c4}) is also
regular.

\section{The special gauge}

\label{specialgauge}

\setcounter{equation}{0}

We have seen that the only covariant gauge that leads to the pseudounitarity
equation is the Feynman gauge, which corresponds to formula (\ref{nocon})
with $\lambda =1$. However, the Feynman gauge has ghosts, that is to say the
matrix $H$ of formula (\ref{puni}) has negative entries.

The Feynman gauges do not make unitarity manifest. Actually, all the
propagators (\ref{zero}) for $n>0$ have ghosts. It is hard to prove that the
ghosts compensate each other in the Feynman gauge, although not impossible 
\cite{thooft,brsunitarity}. Here we prefer to follow a different strategy,
which amounts to prove perturbative unitarity in gauge theories and gravity
by working in a new, noncovariant gauge that satisfies all the requirements
we have outlined and interpolates between the Feynman gauge and the Coulomb
gauge.

We call the new gauge \textquotedblleft special\textquotedblright , because
of its properties. In this section we build the special gauge in Abelian and
non-Abelian gauge theories, while in section \ref{qg} we build it in quantum
gravity. We work in arbitrary dimensions\footnote{%
The case $d=3$, which we do not treat here in detail, can be studied by
including the Chern-Simons term, to avoid the infrared problems due to the
superrenormalizability of the gauge coupling.} $d$ greater than 3. The gauge
group indices are understood in the formulas written below.

Consider the gauge-fixed Lagrangian%
\begin{equation}
\mathcal{L}_{\text{YM}}=-\frac{1}{4}F_{\mu \nu }F^{\mu \nu }-\frac{1}{%
2\lambda }\mathcal{G}^{2}(A)-\bar{C}\mathcal{G}(DC),  \label{lym}
\end{equation}%
where $\mathcal{G}(A)$ is the gauge choice, which is assumed to be linear in 
$A$, while $D$ denotes the covariant derivative. In $d>4$ the theory is
nonrenormalizable, so we should include infinitely many corrections of
higher dimensions, which are optional in $d=4$. We do not write them
explicitly, because, for our purposes, it is sufficient to assume that they
are perturbative, local and Hermitian. We also omit the matter
contributions, which are not important for the moment.

Now, take 
\begin{equation}
\mathcal{G}(A)=\zeta \partial _{0}A_{0}+\mathbf{\nabla }\cdot \mathbf{A,}
\label{gao}
\end{equation}%
where $\zeta $ is another gauge-fixing parameter. For $\zeta =1,0$ we have
the Lorenz and Coulomb gauges, respectively.

The gauge field propagators read 
\begin{eqnarray}
\left\langle A^{0}(k)A^{0}(-k)\right\rangle _{0} &=&-\frac{i(\lambda E^{2}-%
\mathbf{k}^{2})}{(\zeta E^{2}-\mathbf{k}^{2})^{2}},\qquad \left\langle
A^{i}(k)A^{0}(-k)\right\rangle _{0}=\frac{i(\zeta -\lambda )k^{i}E}{(\zeta
E^{2}-\mathbf{k}^{2})^{2}},  \notag \\
\left\langle A^{i}(k)A^{j}(-k)\right\rangle _{0} &=&\frac{i\Pi ^{ij}}{E^{2}-%
\mathbf{k}^{2}}+\frac{i(\zeta ^{2}E^{2}-\lambda \mathbf{k}^{2})}{(\zeta
E^{2}-\mathbf{k}^{2})^{2}}\frac{k^{i}k^{j}}{\mathbf{k}^{2}},  \label{bisa1}
\end{eqnarray}%
where the $\langle \cdots \rangle _{0}$ denotes the free field limit of the
average and%
\begin{equation}
\Pi ^{ij}=\delta ^{ij}-\frac{k^{i}k^{j}}{\mathbf{k}^{2}}  \label{proja}
\end{equation}%
is the projector onto the transversal components of the gauge field. The
ghost propagator is 
\begin{equation}
\left\langle C(k)\bar{C}(-k)\right\rangle _{0}=\frac{i}{\zeta E^{2}-\mathbf{k%
}^{2}}.  \label{bisa2}
\end{equation}

The propagators just listed do not satisfy the assumptions required by the
pseudounitarity equation for generic values of $\lambda $ and $\zeta $,
because they have double poles. The special gauge is defined as the one with 
$\lambda =\zeta >0$, where the double poles disappear. We obtain%
\begin{eqnarray}
\left\langle A^{0}(k)A^{0}(-k)\right\rangle _{0} &=&\frac{-i}{\lambda E^{2}-%
\mathbf{k}^{2}+i\epsilon }=-\left\langle C(k)\bar{C}(-k)\right\rangle
_{0},\qquad \left\langle A^{i}(k)A^{0}(-k)\right\rangle _{0}=0,  \notag \\
\left\langle A^{i}(k)A^{j}(-k)\right\rangle _{0} &=&\frac{i\Pi ^{ij}}{E^{2}-%
\mathbf{k}^{2}+i\epsilon }+\frac{i\lambda }{\lambda E^{2}-\mathbf{k}%
^{2}+i\epsilon }\frac{k^{i}k^{j}}{\mathbf{k}^{2}},  \label{bust}
\end{eqnarray}%
where we have inserted the contour prescriptions, which are now
straightforward. Note that formulas (\ref{bust}) have good power counting
behaviors. In particular, the denominators $\mathbf{k}^{2}$ cancel out in
the sum. The KL spectral representation (\ref{genKL}) is satisfied, although
the densities $\rho $ are not positive definite.

The limit $\lambda \rightarrow 0^{+}$ takes us to Coulomb gauge, actually
the Landau limit of the Coulomb gauge. There, the assumptions we have made
do not hold, because, for example, $\left\langle
A^{0}(x)A^{0}(y)\right\rangle $ is proportional to $\delta (x^{0}-y^{0})$,
which violates (\ref{teta}). In the next section we show that in QED there
is a way to circumvent this difficulty and work directly at $\lambda =0$.
Instead, in non-Abelian gauge theories (and \textit{a fortiori} gravity) it
is necessary to work at $\lambda \neq 0$.

To deal with the infrared divergences, we need to introduce an infrared
cutoff $m_{\gamma }$ and remove it later. This can be done in various ways.
We describe two methods that are good for our purposes, a simpler one and a
more involved one. The simpler method works well in renormalizable theories,
the more involved one is designed to work in nonrenormalizable theories.

\subsection{Renormalizable theories}

For the moment, we concentrate on $d=4$ renormalizable Abelian and
non-Abelian gauge theories, possibly coupled to matter. There, it is
sufficient to replace the propagators (\ref{bust}) with%
\begin{eqnarray}
\left\langle A^{0}(k)A^{0}(-k)\right\rangle _{0} &=&-\frac{i}{\lambda E^{2}-%
\mathbf{k}^{2}-m_{\gamma }^{2}+i\epsilon }=-\left\langle C(k)\bar{C}%
(-k)\right\rangle _{0},\qquad \left\langle A^{i}(k)A^{0}(-k)\right\rangle
_{0}=0,  \notag \\
\left\langle A^{i}(k)A^{j}(-k)\right\rangle _{0} &=&\frac{i\Pi ^{ij}}{E^{2}-%
\mathbf{k}^{2}-m_{\gamma }^{2}+i\epsilon }+\frac{i\lambda }{\lambda E^{2}-%
\mathbf{k}^{2}-m_{\gamma }^{2}+i\epsilon }\frac{k^{i}k^{j}}{\mathbf{k}^{2}}.
\label{specialone}
\end{eqnarray}

The Lagrangian that leads to the propagators (\ref{specialone}) is equal to (%
\ref{lym}) plus the mass terms 
\begin{equation}
\mathcal{L}_{m_{\gamma }}=\frac{m_{\gamma }^{2}}{2}A_{\mu }A^{\mu }+\frac{%
m_{\gamma }^{2}}{2\lambda }(1-\lambda )(\boldsymbol{\nabla }\cdot \mathbf{A)}%
\frac{1}{\Delta }(\boldsymbol{\nabla }\cdot \mathbf{A)}-m_{\gamma }^{2}\bar{C%
}C,  \label{lmg}
\end{equation}%
and is nonlocal in space.

We must show that the regularization defined in subsection \ref{regula} is
well defined in the special gauge, because the denominator $\mathbf{k}^{2}$
of the projector $k^{i}k^{j}/\mathbf{k}^{2}$ does not have the form
specified in formula (\ref{prostru}). We must also pay attention to the
contact terms, because the procedure of subsection \ref{contact} to deal
with them does depend on (\ref{prostru}).

For definiteness, we call an expression \textit{regular} if it just involves
denominators equal to products of polynomials $aE^{2}-b\mathbf{k}%
^{2}-m^{2}+i\epsilon $ with $a>0$ and $b>0$. We call any other expression 
\textit{irregular}. For example, the projector $k^{i}k^{j}/\mathbf{k}^{2}$
is irregular, although it has good infrared and ultraviolet behaviors.

Let us point out a few obvious facts. The propagators of the bosonic fields
and those of the Faddeev-Popov ghosts decrease like $1/E^{2}$ at large
energies. Instead, the fermionic propagators decrease like $1/E$. Call the
vertices that carry at least three legs \textquotedblleft
proper\textquotedblright . In $d=4$ renormalizable Abelian and non-Abelian
gauge theories coupled to matter the proper vertices with no fermionic legs
contain at most one time derivative, while the proper vertices involving
fermionic legs have no time derivatives. Finally, the irregular
contributions to $\left\langle A^{i}(k)A^{j}(-k)\right\rangle _{0}$ read%
\begin{equation}
\frac{i(1-\lambda )m_{\gamma }^{2}}{(\lambda E^{2}-\mathbf{k}^{2}-m_{\gamma
}^{2}+i\epsilon )(E^{2}-\mathbf{k}^{2}-m_{\gamma }^{2}+i\epsilon )}\frac{%
k^{i}k^{j}}{\mathbf{k}^{2}}  \label{pote}
\end{equation}%
and behave like $1/E^{4}$ at large energies.

We treat the quadratic counterterms as two-leg vertices. To minimize the
number of time derivatives acting on the same propagator in coordinate
space, the kinetic counterterms are assumed to have the forms $\sim
(\partial _{0}\phi )^{2}$ for bosons and $\sim \bar{\psi}\partial _{0}\psi $
for fermions.

Now, consider a Feynman diagram. The power of the energy brought by the
vertices of each loop is at most equal to the number of proper vertices with
no fermionic legs, plus twice the number of bosonic quadratic counterterms,
plus the number of fermionic quadratic counterterms. Then, if we ignore the
tadpoles for a moment, every energy integral is convergent (before
integrating on the space momenta) and the multiple energy integrals are
overall convergent.

The tadpoles can be treated apart. The fermionic tadpole is straightforward,
because it does not involve the projector $k^{i}k^{j}/\mathbf{k}^{2}$. The
bosonic tadpole involves it by means of (\ref{pote}), which contributes to a
convergent energy integral. Moreover, adopting the prescription of symmetric
integration, the energy integrals are convergent in both types of tadpoles.

Thus, the regularization of subsection \ref{regula} is well defined. The
integrals on the energy and those on the space momenta can be freely
interchanged.

The same arguments prove that the irregular contributions (\ref{pote}) to
the propagators cannot generate contact terms. Indeed, the vertices cannot
provide enough $E$ powers to compensate the $E^{4}$ appearing in the
denominator of (\ref{pote}). Thus, the contact terms can only come from the
regular terms and can be treated as explained in subsection \ref{contact}.

The locality of counterterms is usually proved by differentiating the
Feynman diagrams with respect to the external energies and space momenta,
then showing that a sufficient number of such derivatives makes the
integrals overall convergent \cite{collins}. This strategy works when the
integrands are regular. Instead, the derivatives of $k^{i}k^{j}/\mathbf{k}%
^{2}$ with respect to the components of $\mathbf{k}$ just improve the
ultraviolet behavior of the integral on $\mathbf{k}$, but do not improve the
behavior of the integral on the energy $E$. Nevertheless, we have shown that
all the integrals on the energies are convergent by themselves, so their
ultraviolet behaviors do not need to be improved. For this reason, a
sufficient number of derivatives with respect to the external energies and
space momenta does make a diagram overall convergent. Once the diagram is
equipped with the counterterms that subtract its subdivergences, the same
operation makes the sum fully convergent. It is easy to check that all the
regions of integration are properly subtracted. This proves the locality of
counterterms. For similar reasons, it is straightforward to prove that the
counterterms are polynomial in $m_{\gamma }^{2}$.

In the end, the renormalization in the special gauge is straightforward. The
renormalized Lagrangian coincides with the one at $m_{\gamma }=0$ plus the
counterterms 
\begin{equation*}
\Delta \mathcal{L}_{m_{\gamma }}=\frac{m_{\gamma }^{2}}{2}\Delta
Z_{0}(A^{0})^{2}-\frac{m_{\gamma }^{2}}{2}\Delta Z_{s}(A^{i})^{2}-m_{\gamma
}^{2}\Delta Z_{g}\bar{C}C,
\end{equation*}%
where $\Delta Z_{0}$, $\Delta Z_{s}$ and $\Delta Z_{g}$ are divergent
constants.

The other requirements of the previous sections are fulfilled, before and
after renormalization. This ensures that the largest time equation, the
cutting equations and the pseudounitarity equation hold in the special
gauge\ for arbitrary $\lambda >0$ in $d=4$, if the theory is renormalizable
by power counting.

\subsection{Nonrenormalizable theories}

The construction just given is sufficient for renormalizable gauge theories,
such as the standard model in flat space. In view of the generalization to
quantum gravity, we explain how to adapt the special gauge to
nonrenormalizable gauge theories in arbitrary dimensions $d>3$.

We introduce two fictitious masses, $\mu $ and $m_{\gamma }$, which play
different roles. Define%
\begin{equation*}
P_{\lambda ,\theta ,\eta }\equiv \frac{1}{\lambda E^{2}-\theta \mathbf{k}%
^{2}-\eta \mu ^{2}-m_{\gamma }^{2}+i\epsilon }
\end{equation*}%
and replace the propagator $\left\langle A^{i}(k)A^{j}(-k)\right\rangle _{0}$
of (\ref{specialone}) by 
\begin{equation}
\left\langle A^{i}(k)A^{j}(-k)\right\rangle _{0}=iP_{1,1,1}\delta
^{ij}+i\left( Q_{N}(\lambda ,r)-P_{1,1,1}\right) \frac{k^{i}k^{j}}{\mathbf{k}%
^{2}+\mu ^{2}},  \label{modaa}
\end{equation}%
where%
\begin{equation}
Q_{N}(\lambda ,r)\equiv \lambda \sum_{n=0}^{N}m_{\gamma }^{2n}(\lambda
-1)^{n}\prod\limits_{q=0}^{n}P_{\lambda ,r_{q},r_{q}},  \label{qn}
\end{equation}%
$r\equiv \{r_{0},r_{1},\ldots \}$ and $r_{q}$ are positive constants such
that $r_{q}\neq r_{q^{\prime }}$ for $q\neq q^{\prime }$, with $r_{0}=1$.
For example, we can choose $r_{q}=q+1$.

Before explaining where the idea for the replacement (\ref{modaa}) comes
from, we give its key properties, in connection with unitarity and
renormalization.

It is easy to prove the identity 
\begin{equation}
Q_{N}(\lambda ,r)=\sum_{n=0}^{N}P_{\lambda ,r_{n},r_{n}}\sum_{q=n}^{N}\left( 
\frac{m_{\gamma }^{2}}{\mathbf{k}^{2}+\mu ^{2}}\right) ^{q}c_{nq}(\lambda ),
\label{sumpo}
\end{equation}%
where $c_{nq}(\lambda )$ are polynomials of $\lambda $. We see that $\mu $
regulates the infrared divergences that would appear in the individual terms
on the right-hand side of this formula at $\mathbf{k}=0$.

The irregular term $ik^{i}k^{j}/(\mathbf{k}^{2}+\mu ^{2})$ inside $%
\left\langle A^{i}(k)A^{j}(-k)\right\rangle _{0}$ is multiplied by the
difference $Q_{N}(\lambda ,r)-P_{1,1,1}$, which satisfies the property 
\begin{equation}
\frac{ik^{i}k^{j}}{\mathbf{k}^{2}+\mu ^{2}}\left( Q_{N}(\lambda
,r)-P_{1,1,1}\right) =-i\frac{k^{i}k^{j}}{\mathbf{k}^{2}+\mu ^{2}}\frac{%
(\lambda -1)^{N+1}m_{\gamma }^{2N+2}}{E^{2}-\mathbf{k}^{2}-\mu
^{2}-m_{\gamma }^{2}+i\epsilon }\prod\limits_{n=0}^{N}P_{\lambda
,r_{n},r_{n}}+\text{ regular terms.}  \label{behav}
\end{equation}%
To derive this formula, note that if we set $r_{q}=1$ for every $q$ and $%
N=\infty $, the function $Q_{N}(\lambda ,r)$ resums into%
\begin{equation}
Q_{\infty }(\lambda )\equiv \lambda \sum_{n=0}^{\infty }m_{\gamma
}^{2n}(\lambda -1)^{n}\prod\limits_{q=0}^{n}P_{\lambda ,1,1}=\frac{\lambda }{%
\lambda E^{2}-\mathbf{k}^{2}-\mu ^{2}-\lambda m_{\gamma }^{2}+i\epsilon }.
\label{resa}
\end{equation}%
Replacing $Q_{N}(\lambda ,r)$ by $Q_{\infty }(\lambda )$ in (\ref{modaa}),
we obtain%
\begin{equation}
\left\langle A^{i}(k)A^{j}(-k)\right\rangle _{0}=iP_{1,1,1}\delta ^{ij}+%
\frac{i(1-\lambda )k^{i}k^{j}}{(E^{2}-\mathbf{k}^{2}-\mu ^{2}-m_{\gamma
}^{2}+i\epsilon )(\lambda E^{2}-\mathbf{k}^{2}-\mu ^{2}-\lambda m_{\gamma
}^{2}+i\epsilon )}.  \label{lmp}
\end{equation}%
Then, there are no irregular terms, the regularization of subsection \ref%
{regula} is well defined, the contact terms are under control by means of
the procedure of subsection \ref{contact} and the locality of counterterms
is obvious. The point is that the arguments of section \ref{nonabe} about
unitarity do not work well with the choice (\ref{lmp}), because (\ref{resa})
shows that the squared mass $m_{\gamma }^{2}$ gets multiplied by $\lambda $
in some cuts, which invalidates the inequality (\ref{impossible}) at $\mu =0$%
.

Nonetheless, the resummation (\ref{resa}) gives us the inspiration for the
replacement (\ref{modaa}). Indeed, truncate the sum of (\ref{resa}) to $%
N<\infty $ and replace the coefficients of $\mathbf{k}^{2}+\mu ^{2}$ in the
denominators with arbitrary numbers $r_{0},r_{1},\ldots $, so as to obtain (%
\ref{qn}). It is clear that these operations lead to the behavior (\ref%
{behav}). In particular, the variations of the $\mathbf{k}^{2}+\mu ^{2}$
coefficients just affect the regular terms. The role of those coefficients
is to make sure that there are no double poles.

Now we use (\ref{behav}) to prove that the modification (\ref{modaa}) has
the properties we need, that is to say the regularization of subsection \ref%
{regula} is well defined, the contact terms are under control and the
locality of counterterms holds. Such properties are obviously satisfied by
the regular contributions to the Feynman diagrams, so we can concentrate on
the contributions that involve irregular terms.

We recall that we are considering a nonrenormalizable theory, whose
Lagrangian contains infinitely many vertices. It is helpful to expand the
interaction Lagrangian in powers of the energy and focus on some finite
truncation. If, at the same time, we truncate the loop expansion to a finite
order, only a finite number of amplitudes, vertices and diagrams are
involved in every calculation. So doing, we are able to prove perturbative
unitarity within any finite truncation, which is enough to prove
perturbative unitarity for the whole theory.

By formula (\ref{behav}), there exists an $N$ such that all the irregular
contributions to the Feynman diagrams are overall convergent within the
truncation. At\ one loop, the integrals that contain irregular terms are
convergent by themselves. At higher orders, they are convergent once the
counterterms that subtract the subdivergences (associated with the regular
contributions to the subdiagrams) are included. Thus, the regularization of
subsection \ref{regula} is well defined. Since the irregular terms do not
contribute to the renormalization of the theory, the locality of
counterterms obviously holds. Moreover, formula (\ref{behav}) shows that for 
$N$ large enough the vertices cannot provide enough powers of $E$ to match
the total $E$ powers appearing in the denominators of the irregular terms.
This means that the contact terms are local, within the truncation, because
they can only be generated by the regular terms. This fact, together with
the locality of the vertices, ensures that the procedure of subsection \ref%
{contact} to deal with the contact terms is still valid.

The construction also works in the case of renormalizable theories, where it
is sufficient to choose $N>d/2-1$.

Formulas (\ref{sumpo}) and (\ref{modaa}) show that the assumptions that lead
to the pseudounitarity equation are satisfied at $\lambda >0$, $m_{\gamma
}\neq 0$, for arbitrary $N\geqslant 0$.

Some remarks are in order, about the recovery of gauge invariance and gauge
independence when $m_{\gamma }$ is sent back to zero. Using the
Batalin-Vilkovisky formalism \cite{bata}, gauge invariance is encoded into
the antiparentheses $(S,S)=2(S,S_{m_{\gamma }})$, where $S_{m_{\gamma
}}=\int \mathcal{L}_{m_{\gamma }}$ collects the $m_{\gamma }$ mass terms,
while gauge independence is encoded in the expression%
\begin{equation*}
\frac{\partial S}{\partial \lambda }-(S,\Psi _{\lambda })=\frac{\partial
S_{m_{\gamma }}}{\partial \lambda },
\end{equation*}%
where $\Psi _{\lambda }$ is the $\lambda $ derivative of the gauge fermion $%
\Psi $, which is the local functional that performs the gauge fixing (for a
recent reference, with details and the notation, see \cite{ABward}). The
right-hand sides of both equations should vanish, at least when $\varepsilon
=0$, but they do not if $m_{\gamma }\neq 0$. Their effects on the generating
functional $\Gamma $ of the one-particle irreducible correlation functions
are encoded into the averages $2\langle (S,S_{m_{\gamma }})\rangle $ and $%
\langle \partial S_{m_{\gamma }}/\partial \lambda \rangle $, which contain
insertions of new vertices besides those of the standard Feynman rules. We
want to make sure that the Feynman diagrams that contain such insertions are
also well regularized and satisfy the locality of counterterms, and check
that their contact terms are still under control.

Write the free massive Lagrangian in compact notation as%
\begin{equation*}
\mathcal{L}_{\text{free}}+\mathcal{L}_{m_{\gamma }}=\frac{1}{2}\Phi ^{\alpha
}Q_{\alpha \beta }\Phi ^{\beta },
\end{equation*}%
where $\Phi ^{\alpha }$ are all the fields (including the ghosts, the
antighosts and the Lagrange multipliers for the gauge fixing). We have%
\begin{equation*}
\frac{\partial S_{m_{\gamma }}}{\partial \lambda }=\frac{1}{2}\int \Phi
^{\alpha }\frac{\partial Q_{\alpha \beta }}{\partial \lambda }\Phi ^{\beta }-%
\frac{\partial }{\partial \lambda }\int \mathcal{L}_{\text{free}}.
\end{equation*}%
The last contribution is local. The other term gives $-(i/2)(\partial
f^{\alpha \beta }/\partial \lambda )$, if we include the propagators $%
f^{\alpha \beta }$ attached to its legs. The irregular part can then be
easily derived from formula (\ref{behav}). Again, if $N$ is large enough
this irregular insertion cannot generate contact terms and every subintegral
that contains it is overall convergent.

Similarly,%
\begin{equation*}
2(S,S_{m_{\gamma }})=(S,\Phi ^{\alpha })Q_{\alpha \beta }\Phi ^{\beta
}-2(S,S_{\text{free}}).
\end{equation*}%
The last contribution is local, while the other term becomes local,
precisely equal to $i(S,\Phi ^{\alpha })$, once it is inserted in a Feynman
diagram and the propagator attached to the right field $\Phi $ is included.

\section{Proof of unitarity in QED}

\label{qed}

\setcounter{equation}{0}

We are now ready to give the simplest proof of perturbative unitarity in
gauge theories, which applies to QED (in $d=4$).

First we show that the Feynman diagrams can be calculated directly in the
Coulomb gauge, which can be reached as the $\lambda \rightarrow 0$ limit of
the special gauge defined in the previous section. The quadratic part of the
Lagrangian is singular at $\lambda =0$, but the Feynman rules are regular.
The propagators are%
\begin{eqnarray}
\left\langle A^{0}(k)A^{0}(-k)\right\rangle _{0} &=&\frac{i}{\mathbf{k}%
^{2}+m_{\gamma }^{2}}=-\left\langle C(k)\bar{C}(-k)\right\rangle _{0},\qquad
\left\langle A^{i}(k)A^{j}(-k)\right\rangle _{0}=\frac{i\Pi ^{ij}}{E^{2}-%
\mathbf{k}^{2}-m_{\gamma }^{2}+i\epsilon },  \notag \\
\left\langle A^{i}(k)A^{0}(-k)\right\rangle _{0} &=&0,\qquad \left\langle
\psi (p)\bar{\psi}(-p)\right\rangle _{0}=\frac{i(p_{\mu }\gamma ^{\mu }+m)}{%
p^{2}-m^{2}+i\epsilon },  \label{trasva}
\end{eqnarray}%
while the vertices are encoded in the interaction Lagrangian $\mathcal{L}%
_{I}=-eA_{\mu }\bar{\psi}\gamma ^{\mu }\psi $.

We want to show that the correlation functions are also regular in the limit 
$\lambda \rightarrow 0$. The problem is that the propagators $\left\langle
A^{0}A^{0}\right\rangle _{0}$ and $\left\langle C\bar{C}\right\rangle _{0}$
do not have the form (\ref{prostru}) at $\lambda =0$. Moreover, the
propagator of the longitudinal component $\mathbf{A}_{\parallel }\equiv (1/%
\sqrt{-\Delta })\boldsymbol{\nabla }\cdot \mathbf{A}$ of the photon naively
disappears in the limit, because it is multiplied by $\lambda $. We must
show that the contributions coming from this component\ can be consistently
dropped from the Feynman diagrams. The ghosts can be ignored, because they
decouple.

Each loop contains at least one different fermion propagator, which makes
the energy integrals behave at worst like $\sim \int \mathrm{d}E/E$ for
large energies $E$. Such integrals are convergent by symmetric integration.
At $\lambda >0$ the symmetric integration is justified by the regularization
of section \ref{s1}. At $\lambda =0$ it must be assumed by default. Then,
the Feynman diagrams are well regularized. Moreover, the energy integrals
can be freely interchanged with the integrals on the space momenta. Finally,
the extra $\lambda $ factor carried by the $\mathbf{A}_{\parallel }$
propagator multiplies integrals that are regular for $\lambda \rightarrow 0$%
. Thus, every diagram that has internal $\mathbf{A}_{\parallel }$ lines
disappears in the limit. The regular behavior for $\lambda \rightarrow 0$
implies that the counterterms of the Coulomb gauge are the $\lambda
\rightarrow 0$ limit of those evaluated at $\lambda >0$. The diagrams that
contain external $\mathbf{A}_{\parallel }$ legs can be ignored, as well as
their counterterms, because, by formula (\ref{trasva}), the vertices that
depend on $\mathbf{A}_{\parallel }$ do not contribute beyond the tree level.

Now we assume that $\lambda $ vanishes and inquire about unitarity. The main
difficulty is that the propagator $\left\langle A^{0}A^{0}\right\rangle _{0}$
does not satisfy the assumptions that lead to the pseudounitarity equation,
because it does not have the form (\ref{genKL}). We can solve this problem
as follows. Since we do not need $A^{0}$ on the external legs of equation (%
\ref{ltefr}), we integrate it out.

Consider a generic (uncut) Feynman diagram with $A^{i}$, $\psi $ and $\bar{%
\psi}$ on the external legs. Every $A^{0}$ internal leg must connect two
vertices proportional to $A_{0}\psi ^{\dagger }\psi $. Focus on a subdiagram
made by such vertices and the $A^{0}$ propagator that connects them. Replace
this subdiagram with an effective four fermion vertex proportional to $\psi
^{\dagger }\psi \left\langle A^{0}A^{\prime \hspace{0.01in}0}\right\rangle
_{0}\psi ^{\prime \hspace{0.01in}\dagger }\psi ^{\prime }$. This operation
is equivalent to replace the Feynman rules listed above with the ones where
the propagators are just $\left\langle A^{i}A^{j}\right\rangle _{0}$ and $%
\left\langle \psi \bar{\psi}\right\rangle _{0}$, while the vertices are
those encoded in the effective interaction Lagrangian 
\begin{equation}
\mathcal{L}_{I\text{eff}}(t,\mathbf{r})=-\frac{m_{\gamma }e^{2}}{8\pi }\int
\rho (t,\mathbf{r})V(m_{\gamma }|\mathbf{r}-\mathbf{r}^{\prime }|)\rho (t,%
\mathbf{r}^{\prime })\mathrm{d}^{3-\varepsilon }\mathbf{r}^{\prime }+e%
\hspace{0.01in}\mathbf{J}(t,\mathbf{r})\cdot \mathbf{A}(t,\mathbf{r})
\label{yuka}
\end{equation}%
(at the tree level), where $\rho =\psi ^{\dagger }\psi $ and $J^{i}=\bar{\psi%
}\gamma ^{i}\psi $, while the function $V(x)$ can be considered as the
dimensional continuation of the Yukawa potential [indeed, $V(x)=\mathrm{e}%
^{-x}/x$ at $\varepsilon =0$]. At the renormalized level, $\mathcal{L}_{I%
\text{eff}}(t,\mathbf{r})$ keeps its form, except for renormalization
constants in front of the two vertices and $m_{\gamma }$.

The new Feynman rules satisfy our assumptions. Observe that (\ref{yuka}) is
local in time and nonlocal in space, which means that assumption ($a$) is
satisfied, but assumption ($a^{\prime }$) is not. This is not a problem,
because assumption ($a^{\prime }$) is just required to handle the loops of
contact terms, which cannot be generated here. Moreover, the propagators $%
\left\langle A^{i}A^{j}\right\rangle _{0}$ and $\left\langle \psi \bar{\psi}%
\right\rangle _{0}$ satisfy the general KL\ decomposition (\ref{genKL}).
Thus, the largest time equation (\ref{lte}) and the cutting equations (\ref%
{ltefr}) hold. Since the cut propagators have only (massive) physical
states, this leads to the unitarity equation (\ref{uni}).

Next, we build the physical amplitudes. We still have $m_{\gamma }$, so the
theory defined by the new Feynman rules is not gauge invariant. If the
external fields are off shell, the left-hand side of the cutting equation
does not have infrared divergences when $m_{\gamma }$ tends to zero. Thus,
the sum of the cut diagrams on the right-hand side is also smooth in the
limit $m_{\gamma }\rightarrow 0$.

When we put the external legs on shell, other infrared divergences appear.
They can be canceled by summing cutting equations associated with diagrams
that have the same types of infrared divergences \cite{infred}.

Finally, we must project the external legs onto gauge invariant states. The
external photons $A^{i}(k)$ can be multiplied by the physical polarizations
or the projectors $\Pi ^{ij}$ of (\ref{proja}). The cut photon legs are
already multiplied by such projectors. The external electron legs and the
cut electron legs do not need a special treatment in the Coulomb gauge,
because an insertion of $\psi $ is equivalent to the insertion of the gauge
invariant operator%
\begin{equation*}
\psi ^{\prime }=\psi \hspace{0.01in}\mathrm{\exp }\left( ie\frac{1}{\Delta }%
\boldsymbol{\nabla }\cdot \mathbf{A}\right) .
\end{equation*}%
The transverse form of the propagator $\left\langle A^{i}A^{j}\right\rangle
_{0}$ appearing in formula (\ref{trasva}) shows that the insertions of $\psi
^{\prime }$ and $\psi $ in the correlation functions give the same results.

The operations just described allow us to conclude that the $S$ matrix of
perturbative QED is unitary, as desired. Since the physical amplitudes are
gauge independent, the conclusion extends from the Coulomb gauge to any
other gauge.

\section{Proof of unitarity in non-Abelian gauge theories}

\label{nonabe}

\setcounter{equation}{0}

In this section we prove the perturbative unitarity of non-Abelian gauge
theories, while in the next section we extend the proof to quantum gravity.

The arguments of the previous section do not generalize beyond QED. Indeed,
in non-Abelian gauge theories and quantum gravity it is not consistent to
work at $\lambda =0$, where several Feynman diagrams become singular. For
example, it is possible to build loops of circulating ghosts $C$, $\bar{C}$
and/or gauge fields $A_{0}$. According to formula (\ref{trasva}), the
integrands of such loops are polynomial in the energies, but not in the
space momenta, so the integrals are ill defined.

If we could dimensionally continue the energies, the mentioned integrals
would vanish. However, since we work in Minkowski spacetime, a continuation
of the energy clashes against assumptions (\ref{teta}) and (\ref{assd}). One
may think of using an \textit{ad hoc} regularization just for the energy
integrals, but the removal of that regulator is very problematic.

We avoid these difficulties by working in the special gauge. That is to say,
we use the propagators (\ref{specialone}), possibly with the modification (%
\ref{modaa}), and keep $\lambda >0$. We know from section \ref{specialgauge}
that the assumptions that lead to the pseudounitarity equation are
satisfied, the cut propagators being $2\theta (E)$ or $2\theta (-E)$ times%
\begin{eqnarray}
\mathrm{Im}\left[ i\left\langle A^{0}A^{0}\right\rangle _{0}\right] &=&-\pi
\delta (\lambda E^{2}-\mathbf{k}^{2}-m_{\gamma }^{2})=-\mathrm{Im}\left[
i\left\langle C\bar{C}\right\rangle _{0}\right] ,\qquad \mathrm{Im}\left[
i\left\langle A^{i}A^{0}\right\rangle _{0}\right] =0,  \notag \\
\mathrm{Im}\left[ i\left\langle A^{i}A^{j}\right\rangle _{0}\right] &=&\pi
\delta (E^{2}-\mathbf{k}^{2}-\mu ^{2}-m_{\gamma }^{2})\left( \delta ^{ij}-%
\frac{k^{i}k^{j}}{\mathbf{k}^{2}+\mu ^{2}}\right)  \label{utto} \\
&&+\pi \sum_{n=0}^{N}\delta (\lambda E^{2}-r_{n}\mathbf{k}^{2}-r_{n}\mu
^{2}-m_{\gamma }^{2})\sum_{q=n}^{N}\left( \frac{m_{\gamma }^{2}}{\mathbf{k}%
^{2}+\mu ^{2}}\right) ^{q}c_{nq}(\lambda )\frac{k^{i}k^{j}}{\mathbf{k}%
^{2}+\mu ^{2}}.  \notag
\end{eqnarray}%
In addition to the physical degrees of freedom, we have unphysical ones,
which are the ghosts $C$ and $\bar{C}$, the temporal component $A_{0}$ of
the gauge field and the longitudinal component $\mathbf{A}_{\parallel }$. To
prove that the pseudounitarity equation turns into the unitarity equation,
we must prove that the unphysical degrees of freedom do not contribute, when
the external legs are all physical.

Consider the cutting equation (\ref{ltefr}) associated with some Feynman
diagram and assume that the external legs are physical. Focus on a cut
diagram $G_{C}$ and observe that the incoming and outgoing legs may be
crossed by the cut or not, as shown in figure (\ref{cut}). Denote the total
energy of the uncut (cut) incoming and outgoing legs by $E_{i}$ ($%
E_{i}^{\prime }$) and $E_{o}$ ($E_{o}^{\prime }$), respectively. Denote the
total energy of the cut legs by $E_{c}$, which is also equal to the total
incoming energy $E_{i}+E_{i}^{\prime }$ and the total outgoing energy $%
E_{o}+E_{o}^{\prime }$.

An internal cut leg may give contributions of three types, with dispersion
relations%
\begin{equation}
DR_{1}:E^{2}=\frac{1}{\lambda }(\mathbf{k}^{2}+m_{\gamma }^{2}),\qquad
DR_{2}:E^{2}=\mathbf{k}^{2}+\mu ^{2}+m_{\gamma }^{2},\qquad DR_{3}:E^{2}=%
\frac{1}{\lambda }(r_{n}\mathbf{k}^{2}+r_{n}\mu ^{2}+m_{\gamma }^{2}).
\label{poles}
\end{equation}%
Write $G_{C}$ as a sum $\sum_{I}G_{C}^{(I)}$, where $G_{C}^{(I)}$ is such
that each internal cut leg contributes by means of one dispersion relation.
Decompose the energy $E_{c}$ of $G_{C}^{(I)}$ as the energy $E_{i}^{\prime
}+E_{o}^{\prime }$ of the external cut legs, plus the energy $E_{\lambda }$
of the internal cut legs that contribute by means of $DR_{1}$ or $DR_{3}$,
plus the energy $E_{2}$ of the internal cut legs that contribute by means of 
$DR_{2}$.

Formulas (\ref{poles}) imply $E_{\lambda }\geqslant m_{\gamma }/\sqrt{%
\lambda }$. Thus, whenever $DR_{1}$ or $DR_{3}$ contribute, the total energy 
$E_{\text{tot}}=E_{i}+E_{i}^{\prime }=E_{o}+E_{o}^{\prime }=E_{c}$ satisfies
the inequality 
\begin{equation}
E_{\text{tot}}\geqslant \frac{m_{\gamma }}{\sqrt{\lambda }}.
\label{impossible}
\end{equation}

Assume that $m_{\gamma }$ is fixed and nonvanishing. The total energy $E_{%
\text{tot}}$ has a given, $\lambda $-independent value. Then, when $\lambda $
is small enough, the condition (\ref{impossible}) cannot be satisfied. This
means that for any diagram with physical external legs, any configuration of
external energies and space momenta and any nonvanishing $m_{\gamma }$, the
dispersion relations $DR_{1}$ and $DR_{3}$ of (\ref{poles})\ cannot
contribute to the cuts of formula (\ref{ltefr}), if $\lambda $ belongs to
the interval $(0,m_{\gamma }^{2}/E_{\text{tot}}^{2})$.

For such values of $\lambda $, we can ignore $DR_{1}$ and $DR_{3}$ and the
cut propagators (\ref{utto}) effectively become%
\begin{eqnarray*}
\mathrm{Im}\left[ i\left\langle A^{0}A^{0}\right\rangle _{0}\right] &=&%
\mathrm{Im}\left[ i\left\langle C\bar{C}\right\rangle _{0}\right] =\mathrm{Im%
}\left[ i\left\langle A^{i}A^{0}\right\rangle _{0}\right] =0, \\
\mathrm{Im}\left[ i\left\langle A^{i}A^{j}\right\rangle _{0}\right] &=&\pi
\delta (E^{2}-\mathbf{k}^{2}-\mu ^{2}-m_{\gamma }^{2})\left( \delta ^{ij}-%
\frac{k^{i}k^{j}}{\mathbf{k}^{2}+\mu ^{2}}\right) .
\end{eqnarray*}%
At this point, we can take the limit $\mu \rightarrow 0$, which gives%
\begin{equation}
\mathrm{Im}\left[ i\left\langle A^{0}A^{0}\right\rangle _{0}\right] =\mathrm{%
Im}\left[ i\left\langle C\bar{C}\right\rangle _{0}\right] =\mathrm{Im}\left[
i\left\langle A^{i}A^{0}\right\rangle _{0}\right] =0,\quad \mathrm{Im}\left[
i\left\langle A^{i}A^{j}\right\rangle _{0}\right] =\pi \delta (E^{2}-\mathbf{%
k}^{2}-m_{\gamma }^{2})\Pi ^{ij}.  \label{physo}
\end{equation}%
The limit $\mu \rightarrow 0$ is regular, not only in the cut propagators
with $\lambda \in (0,m_{\gamma }^{2}/E_{\text{tot}}^{2})$, but also in the
uncut ones (\ref{modaa}) and in the behavior (\ref{behav}), because as long
as $m_{\gamma }\neq 0$ there are no infrared divergences in the individual
diagrams of the cutting equation. Note that we cannot take $\mu \rightarrow
0 $ for generic values of $\lambda $, because the right-hand side of (\ref%
{utto}) would give infrared divergences at $\mathbf{k}=0$.

Next, we take the limit $\epsilon \rightarrow 0$ on the cutting equation,
after grouping the cut diagrams as explained in subsection \ref{infra}, so
that each group is regular in the limit.

Formula (\ref{physo}) shows that, in the end, the unphysical degrees of
freedom disappear from the cuts. This is a good starting point, but not the
end of the story, because we must eventually send the fictitious mass $%
m_{\gamma }$ to zero, which makes the interval $(0,m_{\gamma }^{2}/E_{\text{%
tot}}^{2})$ disappear. To overcome this difficulty, we use the following
trick. Before sending $m_{\gamma }$ to zero, we analytically continue the
cutting equation (\ref{ltefr}) in $\lambda $ from the interval $(0,m_{\gamma
}^{2}/E_{\text{tot}}^{2})$ on the real axis to the complex plane.

Consider an uncut Feynman diagram $G$. At $\mu =0$, the propagators of the
unphysical degrees of freedom have denominators $\lambda E^{2}-r_{n}\mathbf{k%
}^{2}-m_{\gamma }^{2}$, so their poles move away from the real axis during
the analytic continuation from real $\lambda $ to complex $\lambda $. The
continuation is consistent if we deform the integration contours so that the
poles are never crossed. The continued $\lambda $ will be called either $%
\zeta $ or $\zeta ^{\ast }$, leading to the continued diagrams $G_{\zeta }$
and $G{}_{\zeta ^{\ast }}$, respectively.

To better understand what we are doing, write the analytically continued
(renormalized) cutting equation (\ref{ltefr}) in the form 
\begin{equation}
G_{\zeta }+G{}_{\zeta ^{\ast }}^{\dagger }=-\sum_{\text{proper cuttings}}%
\tilde{G}_{\zeta }\mathcal{C}\tilde{G}_{\zeta ^{\ast }}^{\prime \hspace{%
0.01in}\dagger },  \label{ltefrc}
\end{equation}%
where $\tilde{G}_{\zeta }$ and$\mathcal{\ }\tilde{G}_{\zeta ^{\ast
}}^{\prime \hspace{0.01in}\dagger }$ denote the analytic continuations of
the amputated subdiagrams identified by the shadowed and unshadowed sides of
the cuts and $\mathcal{C}$ collects the propagators of the cut lines. The
integration on the momenta of the cut legs is understood.

The diagrams $G{}_{\zeta ^{\ast }}^{\dagger }$ and $\tilde{G}_{\zeta ^{\ast
}}^{\prime \hspace{0.01in}\dagger }$ of the shadowed regions are not the
conjugates of the diagrams $G_{\zeta }$ and $\tilde{G}_{\zeta }^{\prime }$
of the unshadowed regions at $\zeta \neq \lambda $, but the conjugates of
the diagrams $G{}_{\zeta ^{\ast }}$ and $\tilde{G}_{\zeta ^{\ast }}^{\prime
} $. Indeed, the continued equation (\ref{ltefrc}) depends only on $\zeta $
and not on $\zeta ^{\ast }$.

Note that the propagators (\ref{specialone}) do not satisfy the KL\ spectral
representation (\ref{genKL}) for complex $\lambda $. That is why we cannot
obtain equation (\ref{ltefrc}) as a standard cutting equation. We must start
from (\ref{ltefr}), specialize $\lambda $ to the interval $(0,m_{\gamma
}^{2}/E_{\text{tot}}^{2})$, send $\mu $ to zero and then analytically
continue from $\lambda $ to $\zeta $. An obvious, but important fact is that
the cut propagators of $\mathcal{C}$ are independent of $\zeta $, because
they are those of the physical degrees of freedom.

Now we study the properties of the analytic continuation from $\lambda $ to $%
\zeta $. The singularities of a Feynman diagram arise when the contours are
pinched\footnote{%
Strictly speaking, we must first decompose the integral $I$ into a sum $%
\sum_{i}I_{i}$ of integrals $I_{i}$, each of which admits a domain of
convergence $D_{i}$ in the sense of the dimensional regularization. Then, we
can study $I_{i}$ in $D_{i}$ and close the contours of integration on the
energies at infinity. So doing, we see that there are no end point
singularities, but just pinching singularities.} and can be studied by means
of the Landau equations \cite{itzykson}. In our case, the Landau equations
are algebraic and imply that the singularities in the $\zeta $ plane are a
finite number of branch points.

Consider all the diagrams $G_{\zeta }$, $G{}_{\zeta ^{\ast }}$, $\tilde{G}%
_{\zeta }$ and $\tilde{G}_{\zeta ^{\ast }}$ at once and the trajectories
described by the locations of their singularities when $m_{\gamma }$ is
small and tends to zero. There exists a $\bar{m}_{\gamma }$ such that, for
every nonvanishing $m_{\gamma }$ smaller than $\bar{m}_{\gamma }$ the
following three facts hold: ($i$) the trajectories are continuous; ($ii$)
those which do not coincide do not intersect; and ($iii$) there exists a
positive $\lambda _{\min }\leqslant m_{\gamma }^{2}/E_{\text{tot}}^{2}$ such
that no singularities occur in the interval $(0,\lambda _{\min })$ on the
real axis.

Define the domain $U$ as the complex plane $\mathbb{C}$ minus the
trajectories with $m_{\gamma }\in (0,\bar{m}_{\gamma })$. Consider a simply
connected open subset $V$ of $U$ (we can take a disk, for simplicity). For
any $m_{\gamma }\in (0,\bar{m}_{\gamma })$, it is possible to analytically
continue the cutting equation (\ref{ltefr}) from $(0,\lambda _{\min })$ to $%
V $. To achieve this goal, it is sufficient to identify a path $\gamma $
that connects the interval $(0,\lambda _{\min })$ to some $z\in V$ and does
not cross any singularity. When $m_{\gamma }$ decreases, the path $\gamma $
can be continuously deformed to avoid the singularities. So doing, the
continued equation (\ref{ltefrc}) holds in $V$ at $m_{\gamma }=0$.

Now we come to the physical amplitudes. We use gauge independence, following
the recent treatment of ref. \cite{ABward}, which covers all gauge theories,
including the nonrenormalizable ones and the potentially anomalous ones (as
long as they are nonanomalous at one loop, by explicit cancellation, and at
higher orders, by the Adler-Bardeen theorem \cite{adlerbardeen}). The
Hermitian conjugates are never involved in the arguments about gauge
independence, so the results of \cite{ABward} also apply to a complex $\zeta 
$.

Start again from $m_{\gamma }\in (0,\bar{m}_{\gamma })$. Pick any
correlation function, make the analytic continuation from $\lambda $ to $%
\zeta $ and identify the domain $V$. At nonzero $m_{\gamma }$, the equations
of gauge dependence are violated. It is easy to show that the violations,
which are due to insertions proportional to $m_{\gamma }^{2}$ [see e.g (\ref%
{lmg})], disappear when $m_{\gamma }$ tends to zero. Consider an (off shell)
Feynman diagram that contains such insertions. The limit $m_{\gamma
}\rightarrow 0$ can generate divergences, but those divergences are always
beaten by the multiplying factors of $m_{\gamma }^{2}$. For example, a
single insertion may generate a logarithmic divergence $\sim \ln m_{\gamma
}^{2}$ in $d=4$, a powerlike divergence $\sim 1/m_{\gamma }$ in $d=3$ and no
singularity in $d>4$. Two insertions may give a powerlike divergence $\sim
m_{\gamma }^{d-6}$ in $d<6$, a logarithmic one $\sim \ln m_{\gamma }^{2}$ in 
$d=6$ and no singularity in $d>6$, and so on. In the end, the violations are
at worst multiplied by $m_{\gamma }^{d-2}\ln m_{\gamma }^{2}$ and $m_{\gamma
}^{d-2}$, or products and powers of these expressions, and so vanish for $%
m_{\gamma }\rightarrow 0$. These properties are true even when we include
the modifications of formula (\ref{modaa}).

Thus, the equations of gauge dependence are satisfied after the limit $%
m_{\gamma }\rightarrow 0$ in $V$. They ensure \cite{ABward} that the $\zeta $
dependence of the generating functional $\Gamma (\Phi ,K,\zeta )$ of the
one-particle irreducible correlation functions can be absorbed into a
canonical transformation $F_{\zeta }$ (in the Batalin-Vilkovisky sense \cite%
{bata}) of the fields $\Phi $ and the sources $K$ coupled to the $\Phi $
transformations, plus redefinitions of the couplings and the other
parameters of $\Gamma $.

Finally, we put the external legs on shell. The infrared divergences
generated by this operation are of a few universal types. The same types
occur in different diagrams, so it is possible to identify combinations of
cutting equations that are free of such divergences. In the rest of the
discussion, we focus on such combinations, which have a well-defined on
shell limit.

As said, the $\zeta $ dependence is encoded into a canonical transformation,
plus redefinitions of the parameters. A canonical transformation maps a
correlation function, evaluated up to some order within the truncation, onto
a sum of correlation functions, each of which satisfies its own cutting
equations (\ref{ltefrc}). What is important for us is that the canonical
transformation is trivial on the legs that are on shell. Indeed, the
nonlinear terms of the field redefinition correspond to insertions of local
composite fields. Once the diagram is amputated, those insertions get
multiplied by inverse propagators, which vanish on shell.

In the end, the combinations of amplitudes that have a well-defined on shell
limit are also independent of $\zeta $ (possibly after redefining the
couplings and the other parameters). In particular, they can be trivially
continued back from $\zeta \in V$ to real positive values $\lambda $.
Moreover, by our construction, they satisfy the unitarity equation (\ref{uni}%
). This concludes the proof.

\subsection{Examples}

The argument involving the analytic continuation from $\lambda $ to $\zeta $
is rather new, at least to our knowledge, so we give some examples in four
dimensions to better visualize what happens.

The simplest situation is the one-loop self-energy of the four-dimensional $%
\varphi ^{3}$ theory with propagator $iP_{\lambda ,1,0}$ and coupling $g$.
The diagram has the usual value apart from a rescaling of the energy by a
factor $1/\sqrt{\lambda }$. Using the Feynman parameters, without including
the combinatorial factor, we get%
\begin{equation*}
\frac{ig^{2}\Gamma \left( \frac{\varepsilon }{2}\right) }{(4\pi
)^{(4-\varepsilon )/2}\sqrt{\lambda }}\int_{0}^{1}\mathrm{d}x\left[ x(1-x)(%
\mathbf{k}^{2}-\lambda E^{2})+m_{\gamma }^{2}-i\epsilon \right]
^{-\varepsilon /2}.
\end{equation*}%
The analytic continuation in $\lambda $ is done from the interval $%
(0,m_{\gamma }^{2}/E^{2})$, where the right-hand side of (\ref{ltefrc})
vanishes, because no physical degrees of freedom are present. The left-hand
side of formula (\ref{ltefrc}) is then 
\begin{equation}
\frac{ig^{2}}{(4\pi )^{2}\sqrt{\zeta }}\int_{0}^{1}\mathrm{d}x\ln \frac{%
x(1-x)(\mathbf{k}^{2}-\zeta E^{2})+m_{\gamma }^{2}+i\epsilon }{x(1-x)(%
\mathbf{k}^{2}-\zeta E^{2})+m_{\gamma }^{2}-i\epsilon },  \label{vanish}
\end{equation}%
where we have taken the limit $\varepsilon \rightarrow 0$, for simplicity.
Thus, the cutting equations imply that formula (\ref{vanish}) should just
give zero.

To prove this fact, it is sufficient to study the analytic continuation of
the function%
\begin{equation}
\Theta (z)\equiv \frac{1}{2\pi i}\ln \frac{z+i\epsilon }{z-i\epsilon }.
\label{cteta}
\end{equation}%
Clearly, $\Theta (z)$ vanishes for $z$ real and positive, so the analytic
continuation of $\Theta (z)$ from the positive real axis gives zero
everywhere. Instead, the analytic continuation from the negative real axis
gives $\Theta =1$ everywhere. In some sense, the function $\Theta (z)$ is
the complex $\theta $ function.

In the end, formula (\ref{vanish}) turns into 
\begin{equation*}
-\frac{g^{2}}{8\pi \sqrt{\zeta }}\int_{0}^{1}\mathrm{d}x\hspace{0.01in}%
\hspace{0.01in}\Theta (a(x)-\zeta ),\qquad
\end{equation*}%
where%
\begin{equation*}
a(x)=\frac{\mathbf{k}^{2}}{E^{2}}+\frac{m_{\gamma }^{2}}{E^{2}x(1-x)}>\frac{%
m_{\gamma }^{2}}{E^{2}}.
\end{equation*}%
The analytic continuation from the interval $(0,m^{2}/E^{2})$ of the real
axis gives zero everywhere, as expected.

To show that the essential features just described survive when the $\lambda 
$ is not just a simple rescaling of the energies, we consider a second
example, which is a variant of the first one where the diagram remains the
same and one internal leg keeps the propagator $iP_{\lambda ,1,0}$, but the
other internal leg gets the propagator $iP_{1,1,0}$. The right-hand side of (%
\ref{ltefrc}) still vanishes for $\lambda <m_{\gamma }^{2}/E^{2}$.

For simplicity, we take a vanishing external space momentum. We find that
formula (\ref{vanish}) is replaced by 
\begin{equation*}
\frac{ig^{2}}{(4\pi )^{2}}\int_{0}^{1}\frac{\mathrm{d}x}{\sqrt{x+\zeta (1-x)}%
}\ln \frac{\left( x+\zeta (1-x)\right) m_{\gamma }^{2}-\zeta
E^{2}x(1-x)+i\epsilon }{\left( x+\zeta (1-x)\right) m_{\gamma }^{2}-\zeta
E^{2}x(1-x)-i\epsilon }.
\end{equation*}%
Again, the analytic continuation from $(0,m_{\gamma }^{2}/E^{2})$ gives zero
on the entire complex plane.

The third example is a one-loop three-point function, where two propagators
are $iP_{1,1,0}$ and one is $iP_{\lambda ,1,0}$, so that the right-hand side
of (\ref{ltefrc}) is nonzero. There is just one nontrivial cut diagram for $%
\lambda <m_{\gamma }^{2}/E_{\text{tot}}^{2}$, which is 
\begin{equation*}
\includegraphics[width=4truecm]{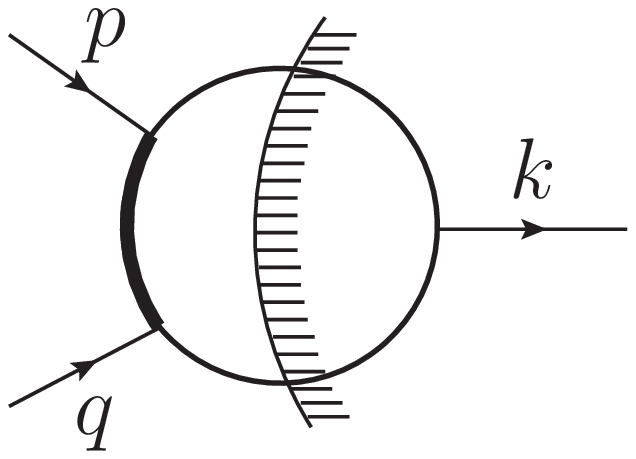}
\end{equation*}%
where the arrows denote the energy flows and the solid line stands for the $%
\lambda $-dependent propagator. To evaluate this diagram, we assume that the
external particle of momentum $k$ is at rest and has mass $M$. We make no
assumptions on the masses of the other external particles. Then we get%
\begin{equation*}
\frac{g^{3}\theta \left( M-2m_{\gamma }\right) }{16\pi M|\mathbf{p}|}\ln 
\frac{4m_{\gamma }^{2}+\left( M\sigma -2|\mathbf{p}|\right) ^{2}-\zeta
\left( M-2p^{0}\right) ^{2}-i\epsilon }{4m_{\gamma }^{2}+\left( M\sigma +2|%
\mathbf{p}|\right) ^{2}-\zeta \left( M-2p^{0}\right) ^{2}-i\epsilon }
\end{equation*}%
where $\sigma =\sqrt{1-\frac{4m_{\gamma }^{2}}{M^{2}}}$. We see that the
analytic continuation from $\lambda $ to $\zeta $ is straightforward, as
well as the limit $m_{\gamma }\rightarrow 0$.

These examples show that no big surprise occurs when we make the operations
described in this section.

We emphasize that the proof we have given, being purely perturbative, does
not deal with nonperturbative issues, such as confinement and chiral
symmetry breaking, which affect the physical spectrum. It simply shows that
the physical amplitudes that are built with the elementary fields satisfy
the unitary cutting equation. The definition of asymptotic states is not
necessary to make these statements meaningful. The true asymptotic states
may be completely different from those suggested by the classical
Lagrangian, as we know from QCD. We recall that, instead of dealing with the
elementary fields, it is often convenient to study the correlation functions
of gauge invariant composite fields and extract the $S$-matrix elements as
residues of their poles \cite{polology}.

\section{ Proof of unitarity in quantum gravity}

\label{qg}

\setcounter{equation}{0}

In this section we generalize the proof of perturbative unitarity to quantum
gravity (with vanishing cosmological constant).

First, we prove that quantum gravity also admits the special gauge, in
arbitrary dimensions $d>3$. The gauge-fixed Lagrangian is%
\begin{equation}
\mathcal{L}_{\text{gf}}=-\frac{1}{2\kappa ^{d-2}}\sqrt{|g|}R+\frac{1}{%
4\lambda _{1}\kappa ^{d-2}}\mathcal{G}_{0}^{2}(g)-\frac{1}{4\lambda
_{2}\kappa ^{d-2}}\mathcal{G}_{i}^{2}(g)+\bar{C}_{0}\mathcal{G}_{0}(%
\overline{DC})-\bar{C}_{i}\mathcal{G}_{i}(\overline{DC}),  \label{lgf}
\end{equation}%
where $\kappa $ is a constant of dimension $-1$ in units of mass, $C_{\mu }$
and $\bar{C}_{\mu }$ are the ghosts and antighosts, respectively, the
gauge-fixing functions $\mathcal{G}_{0}(g)$ and $\mathcal{G}_{i}(g)$ are
linear in the metric tensor $g_{\mu \nu }$ and $\overline{DC}$ stands for $%
D_{\mu }C_{\nu }+D_{\nu }C_{\mu }$.

We start from the most general linear gauge-fixing functions with one
derivative, which are%
\begin{equation*}
\mathcal{G}_{0}(g)=\alpha \partial _{0}g_{ii}+\beta \partial _{i}g_{0i}+\chi
\partial _{0}g_{00},\qquad \mathcal{G}_{i}(g)=\gamma \partial
_{0}g_{0i}+\delta \partial _{i}g_{00}+\xi \partial _{i}g_{jj}+\tau \partial
_{j}g_{ij},
\end{equation*}%
and determine the constants in front of the various terms as follows. First,
we require that the propagators with an odd number of indices 0 vanish.
Second, we simplify the double poles. Third, we eliminate the redundant
constants and arrange the result in the most economic form. At the end, we
find%
\begin{equation*}
\mathcal{G}_{0}(g)=\frac{\lambda }{2}\partial _{0}g_{00}+\frac{1}{2}\partial
_{0}g_{ii}-\partial _{i}g_{0i},\qquad \mathcal{G}_{i}(g)=-\lambda
_{1}\partial _{j}g_{ij}+\frac{1}{2}\left( 2\lambda _{1}-1\right) \partial
_{i}g_{jj}+\lambda \partial _{0}g_{0i}-\frac{\lambda }{2}\partial _{i}g_{00},
\end{equation*}%
together with%
\begin{equation*}
\lambda _{1}=\frac{\lambda (d-3)+d-1}{2(d-2)},\qquad \lambda _{2}=\lambda
\lambda _{1}.
\end{equation*}%
Only one gauge-fixing parameter, which we still call $\lambda $, does
survive.

The ghost propagators are%
\begin{equation}
\langle C^{0}\bar{C}^{0}\rangle _{0}=-i\bar{P}_{\lambda ,1},\qquad \langle
C^{0}\bar{C}^{i}\rangle _{0}=\langle C^{i}\bar{C}^{0}\rangle _{0}=0,\qquad
\langle C^{i}\bar{C}^{j}\rangle _{0}=i\bar{P}_{\lambda ,\lambda _{1}}\Pi
^{ij}+i\bar{P}_{\lambda ,1}\frac{k^{i}k^{j}}{\mathbf{k}^{2}},  \label{pgh}
\end{equation}%
where $\bar{P}_{a,b}=1/(aE^{2}-b\mathbf{k}^{2}+i\epsilon )$. The propagators
of the fluctuations $h_{\mu \nu }=\kappa ^{1-(d/2)}(g_{\mu \nu }-\eta _{\mu
\nu })/2$ around flat space are%
\begin{eqnarray}
\langle h_{00}h_{00}\rangle _{0} &=&\frac{d-3}{d-2}i\bar{P}_{\lambda
,1},\qquad \langle h_{00}h_{ij}\rangle _{0}=\frac{\delta _{ij}}{d-2}i\bar{P}%
_{\lambda ,1},  \notag \\
\langle h_{0i}h_{0j}\rangle _{0} &=&-\frac{i\lambda _{1}}{2}\left( \bar{P}%
_{\lambda ,\lambda _{1}}\Pi _{ij}+\bar{P}_{\lambda ,1}\frac{k_{i}k_{j}}{%
\mathbf{k}^{2}}\right) ,\qquad \langle h_{00}h_{0i}\rangle _{0}=\langle
h_{0i}h_{jk}\rangle _{0}=0,  \label{propag} \\
\langle h_{ij}h_{mn}\rangle _{0} &=&\frac{i\bar{P}_{1,1}}{2}\left( \Pi
_{im}\Pi _{jn}+\Pi _{in}\Pi _{jm}-\frac{2}{d-2}\Pi _{ij}\Pi _{mn}\right) -%
\frac{\lambda }{\mathbf{k}^{2}}\frac{i\bar{P}_{\lambda ,1}}{d-2}\left( \Pi
_{ij}k_{m}k_{n}+k_{i}k_{j}\Pi _{mn}\right)  \notag \\
&&\!\!\!\!\!{+\frac{\lambda i\bar{P}_{\lambda ,\lambda _{1}}}{2\mathbf{k}^{2}%
}\left( \Pi _{im}k_{j}k_{n}+\Pi _{in}k_{j}k_{m}+\Pi _{jm}k_{i}k_{n}+\Pi
_{jn}k_{i}k_{m}\right) +\lambda i\bar{P}_{\lambda ,1}\frac{d-3}{d-2}\frac{%
k_{i}k_{j}k_{m}k_{n}}{(\mathbf{k}^{2})^{2}}.}  \notag
\end{eqnarray}%
At $\lambda =1$ the special gauge coincides with the de Donder one, which is
the gravitational analogue of the Feynman gauge of Yang-Mills theories. In
the limit $\lambda \rightarrow 0$ we get an analogue of the Coulomb gauge,
different from the Prentki gauge.

To have control on the infrared divergences of the individual cut diagrams,
we keep $\langle C^{0}\bar{C}^{i}\rangle _{0}=\langle C^{i}\bar{C}%
^{0}\rangle _{0}=\langle h_{00}h_{0i}\rangle _{0}=\langle
h_{0i}h_{jk}\rangle _{0}=0$ and replace the other propagators with%
\begin{eqnarray*}
\langle C^{0}\bar{C}^{0}\rangle _{0} &=&-iP_{\lambda ,1,1}\hspace{0.01in}%
,\qquad \langle h_{00}h_{00}\rangle _{0}=\frac{d-3}{d-2}iP_{\lambda ,1,1}%
\hspace{0.01in},\qquad \langle h_{00}h_{ij}\rangle _{0}=\frac{\delta _{ij}}{%
d-2}iP_{\lambda ,1,1}\hspace{0.01in}, \\
\langle h_{0i}h_{0j}\rangle _{0} &=&-\frac{i\lambda _{1}}{2\lambda }\left(
Q_{N}(\lambda ,s)\pi _{ij}+Q_{N}(\lambda ,r)\omega _{ij}\right) =-\frac{%
\lambda _{1}}{2}\langle C^{i}\bar{C}^{j}\rangle _{0}, \\
\langle h_{ij}h_{mn}\rangle _{0} &=&\frac{iP_{1,1,1}}{2}\left( \pi _{im}\pi
_{jn}+\pi _{in}\pi _{jm}-\frac{2}{d-2}\pi _{ij}\pi _{mn}\right) -\frac{%
iQ_{N}(\lambda ,r)}{d-2}\left( \pi _{ij}\omega _{mn}+\omega _{ij}\pi
_{mn}\right) \\
&&\!\!\!\!\!\!\!\!{+\frac{i}{2}Q_{N}(\lambda ,s)\left( \pi _{im}\omega
_{jn}+\pi _{in}\omega _{jm}+\pi _{jm}\omega _{in}+\pi _{jn}\omega
_{im}\right) +iQ_{N}(\lambda ,r)\frac{d-3}{d-2}\omega _{ij}\omega _{mn},}
\end{eqnarray*}%
where 
\begin{equation*}
\pi _{ij}=\delta _{ij}-\frac{k_{i}k_{j}}{\mathbf{k}^{2}+\mu ^{2}},\qquad
\omega _{ij}=\frac{k_{i}k_{j}}{\mathbf{k}^{2}+\mu ^{2}},
\end{equation*}%
and the sequence $s=\{s_{0},s_{1},\ldots \}$ is related to $%
r=\{r_{0},r_{1},\ldots \}$ by the formula%
\begin{equation}
s_{n}=\frac{\lambda (d-3)+r_{n}(d-1)}{2(d-2)}.  \label{sr}
\end{equation}%
Moreover, we choose $r_{n}$ such that $r_{n}\neq s_{n^{\prime }}$ for every $%
n$ and $n^{\prime }$, at $\lambda $ small.

As in Yang-Mills theories, the propagators contain irregular terms. It can
be shown that those terms satisfy a property analogous to (\ref{behav}). Precisely,
they are equal to the product of projectors built with $\delta _{ij}$ and $%
\omega _{ij}$ times regular terms that factorize $N+1$ powers of $m_{\gamma
}^{2}$. Note that the irregular denominators can now be as bad as $(\mathbf{k%
}^{2}+\mu ^{2})^{2}$. The relation (\ref{sr}) is crucial to cancel those, up
to $\mathcal{O}(m_{\gamma }^{2N+2})$.

The powers $m_{\gamma }^{2N+2}$ that factorize in front of the irregular
terms lower the degree of divergence. If $N$ is large enough, the irregular
contributions to the Feynman diagrams are overall convergent within any
given truncation. Thus, the locality of counterterms holds within the
truncation, for the same reasons explained in the study of nonrenormalizable
gauge theories. Moreover, for $N$ sufficiently large the irregular parts of
the propagators cannot generate contact terms. Then, contact terms can only
come from the regular contributions and can be dealt with by means of the
procedure explained in subsection \ref{contact}.

The assumptions that are required to derive the pseudounitarity equation are
satisfied at $m_{\gamma }\neq 0$, $\lambda >0$. When $\lambda $ is
sufficiently small, the threshold for the production of the unphysical
particles in the cuts is raised enough to get rid of all of them (once $\mu $
is sent to zero), for any given $m_{\gamma }$ and any total energy of the
incoming particles.

The operators projecting onto the physical degrees of freedom that propagate
in the cuts can be read from formulas (\ref{pgh}) and (\ref{propag}), which
show that $C^{\mu }$, $\bar{C}^{\mu }$, $h_{00}$, $h_{0i}$, $h_{ii}$ and $%
\partial _{j}h_{ij}$ do not propagate for $\lambda $ small. The external
legs can be projected in a similar way. Precisely, it is sufficient to set
the sources coupled to the external legs $C^{\mu }$, $\bar{C}^{\mu }$, $%
h_{00}$ and $h_{0i}$ to zero and project the external legs $h_{ij}$ by means
of 
\begin{equation*}
\left. \prod \right. _{ij,mn}\equiv \frac{1}{2}\left( \Pi _{im}\Pi _{jn}+\Pi
_{in}\Pi _{jm}-\frac{2}{d-2}\Pi _{ij}\Pi _{mn}\right) .
\end{equation*}%
Every other argument of the previous section can be generalized
straightforwardly.

\section{Conclusions}

\label{conclusions}

\setcounter{equation}{0}

In this paper we worked out a proof of perturbative unitarity that is more
economical and general than the ones available in the literature and
applies\ to renormalizable and nonrenormalizable gauge theories and quantum
gravity, in arbitrary dimensions $d$ greater than 3. With an eye on future
generalizations, we searched for the minimum assumptions that lead to the
various equations involved in the proof, which are the largest time
equation, the cutting equations and the pseudounitarity equation. The
minimum assumptions are actually very restrictive and imply a general K\"{a}%
ll\'{e}n-Lehman spectral representation, even if it is not assumed from the
start. The pseudounitarity equation turns into the unitarity equation when
the incoming and outgoing particles, as well as the particles propagating in
the cuts, can be projected onto a subspace of physical states.

We also filled some gaps that exist in the current literature, as in the
treatment of contact terms.

The simplest proof of perturbative unitarity is available in QED by working
directly in the Coulomb gauge. Unfortunately, a similar strategy cannot be
pursued in non-Abelian gauge theories and gravity. In those cases, we
identified a special gauge that fulfills all the assumptions and has several
other virtues. It depends on a unique gauge-fixing parameter $\lambda $ and
interpolates between the Feynman gauge ($\lambda =1$) and the Coulomb gauge (%
$\lambda =0$). When the gauge fields are given fictitious masses to regulate
the on shell infrared divergences, the threshold for the production of the
unphysical particles in the cuts grows while $\lambda $ becomes small.
Eventually, it projects the unphysical particles away. Thus, there exists an
interval of values of $\lambda $ where the pseudounitarity equation turns
into the unitarity equation. To recover gauge invariance, the fictitious
masses must be removed. This can be achieved without jeopardizing the
unitarity equation by making the analytic continuation in $\lambda $.

Various theories do not obey the assumptions that lead to the cutting
equations. Examples are the local higher-derivative theories whose
propagators have poles outside the real axis. Other examples are the
nonlocal theories of gauge fields and gravity formulated in refs. \cite%
{kuzmin}. Indeed, if the vertices are not localized in time, the largest
time equation cannot be derived, because no \textquotedblleft largest
time\textquotedblright\ can be identified in the analogue of the raw diagram 
$F(x_{1}^{0},\cdots ,x_{n}^{0})$. Moreover, if the vertices are nonlocal in
space, the contact terms of subsection \ref{contact} cannot be treated as
explained there. For these reasons, the consistency of the theories of refs. 
\cite{kuzmin} remains an open problem, even if their propagators have no
poles on the complex plane besides the graviton one.

These remarks also suggest an unforseen connection between unitarity and
locality that is worth further investigation.

\vskip 12truept \noindent {\large \textbf{Acknowledgments}}

\vskip 2truept

I am grateful to U. Aglietti, M. Bochicchio, R. Ferrari, M. Mintchev, G.
Paffuti and M. Piva for helpful discussions.

\end{document}